\documentclass[11pt]{article}
\usepackage[margin=0.95in]{geometry}
\usepackage{epsfig,amssymb,soul}

\usepackage{graphicx} 
\usepackage{epstopdf}
\usepackage{amsmath}
\usepackage{amsfonts,amsmath}
\usepackage{bm}
\usepackage{setspace}
\usepackage{comment}
\usepackage{siunitx}
\usepackage{mathrsfs}
\usepackage{multirow}
\usepackage{subcaption}
\usepackage[hang,flushmargin,bottom]{footmisc}
\usepackage{mathtools}
\usepackage{lipsum}
\usepackage{array}
\usepackage{enumerate}
\usepackage{ifthen}
\usepackage{cite}
\usepackage{authblk}
\usepackage{leftidx}
\usepackage{relsize}
\usepackage{braket}
\usepackage{float}
\usepackage{algorithm2e}
\RestyleAlgo{ruled}

\usepackage[autostyle]{csquotes}
\usepackage[dvipsnames]{xcolor}
\newcommand{\etal}{\textit{et al}.~}

\providecommand{\keywords}[1]{\textbf{\textit{Keywords---}} #1}

\title{Reinforcement learning framework for the mechanical design of microelectronic components under multiphysics constraints}
\date{}

\author[1]{Siddharth Nair}
\author[2]{Timothy F. Walsh}
\author[2]{Greg Pickrell}
\author{Fabio Semperlotti \thanks{To whom correspondence should be addressed.\\ \indent\indent Email addresses: nair40@purdue.edu (S. Nair), fsemperl@purdue.edu (F. Semperlotti) }}
\affil[1]{Ray W. Herrick Laboratories, School of Mechanical Engineering,  Purdue University, West Lafayette, IN 47907, USA}
\affil[2]{Sandia National Laboratories, Albuquerque, NM 87185, USA}
\setstcolor{red}

\usepackage{lineno}

\begin{document}
\maketitle


\begin{abstract}

This study focuses on the development of reinforcement learning based techniques for the design of microelectronic components under multiphysics constraints. While traditional design approaches based on global optimization approaches are effective when dealing with a small number of design parameters, as the complexity of the solution space and of the constraints increases different techniques are needed. This is an important reason that makes the design and optimization of microelectronic components (characterized by large solution space and multiphysics constraints) very challenging for traditional methods.
By taking as prototypical elements an application-specific integrated circuit (ASIC) and a heterogeneously integrated (HI) interposer, we develop and numerically test an optimization framework based on reinforcement learning (RL). More specifically, we consider the optimization of the bonded interconnect geometry for an ASIC chip as well as the placement of components on a HI interposer while satisfying thermoelastic and design constraints. This placement problem is particularly interesting because it features a high-dimensional solution space.

\noindent\keywords{Reinforcement learning, Multiphysics constraints, Interconnect design, Heterogeneously integrated (HI) interposer, Q-learning, Proximal policy optimization (PPO)}

\end{abstract}

\section{Introduction}
\label{ssec: Introduction}

The identification of optimal design parameters, shape, and topology of a system to achieve a target physical response is a prototypical class of inverse design problems prevalent across various fields of science and engineering, including but not limited to, solid mechanics \cite{sigmund1996some, maute2013topology}, heat transfer \cite{orlande2012inverse}, acoustics \cite{duhring2008acoustic}, and fluid mechanics \cite{cotter2009bayesian}. While the use of design optimization approaches to address linear and single physics problems is widespread \cite{bendsoe2013topology}, real-world applications often involve coupled, nonlinear, and multiphysics problems. Some examples include the design of microelectronics \cite{sigmund2001design}, discovering mechanical metamaterials with functional characteristics \cite{brown2023deep}, geometry evolution in additive manufacturing \cite{liu2018current}, and development of optical devices in nanophotonics \cite{molesky2018inverse}, to name a few.

Depending on the specific goal of the design, the optimization problem can be formulated from a different perspective focusing, for example, on the topology, shape, or placement aspect. Topology optimization focuses on optimizing material distribution within a design space, while shape optimization involves modifying the geometric boundaries. Further, placement optimization determines the optimal positions of the basic components of an assembly within a given design space.
Topology optimization is traditionally approached using density-based methods like solid isotropic material with penalization (SIMP) technique for material distribution \cite{bendsoe2013topology}, level set methods \cite{wang2003level}, or evolutionary strategies like genetic algorithm (GA) \cite{ohsaki1995genetic}. On the other hand, shape optimization typically employs parameterization methods like non-uniform rational B-splines (NURBS) and is optimized using gradient-based approaches \cite{najafi2015gradient}, adjoint methods \cite{papadimitriou2008aerodynamic}, or quasi-Newton techniques \cite{kunvstek2022quasi}. Both topology and shape optimization can involve a number of design variables, are commonly parameterized, and have been applied to multiphysics problems. However, placement optimization presents different challenges. This approach relies on the ability of global optimization techniques like genetic algorithms (GA) \cite{jung2015sensor}, simulated annealing (SA) \cite{sechen2012vlsi} or particle swarm optimization (PSO) \cite{he2004improved} to solve combinatorial problems. Additionally, placement optimization becomes even more challenging in dynamic design spaces requiring sequential decision-making, where each placement decision can impact subsequent actions and physical characteristics. Such problems are challenging to address using traditional global optimization approaches for multiple reasons. Global optimization approaches typically focus on static objective functions and are not equipped to handle problems involving dynamic design spaces that require sequential decision-making. Moreover, these algorithms generally do not account for temporal dependencies between decisions, making it difficult to attribute future success to earlier decisions. Additionally, global optimization schemes struggle with the \textit{curse of dimensionality} \cite{qian2016derivative}, where the vastness of the solution space and complex interdependencies between design variables increase the likelihood of getting trapped in local minima.

More recently, deep learning based inverse design approaches have gained popularity due to their ability to capture high-dimensional and nonlinear mapping between design parameters and system response. The most common approach involves supervised deep neural network (DNN) models, which have been applied across various computational mechanics design applications, including structures \cite{zhang2021multi, white2019multiscale}, photonics \cite{so2020deep}, acoustics \cite{nair2023grids, wu2022physics}, and functional materials design \cite{guo2021artificial, patnaik2022variable}. Additionally, unsupervised learning techniques, particularly generative DNN models \cite{liu2018generative, so2019designing}, have been widely employed to learn the relationship between inverse designs and their corresponding target functionalities with unlabeled data. 
However, significant amounts of data on the designs of topology, shape, and placement must be provided \textit{a priori} to train these DNNs for accurate predictions. For combinatorial problems like placement optimization, the required volume of data becomes a significant challenge as no dataset is comprehensive enough to capture all possible scenarios, and it is difficult to determine the optimal data distribution necessary to find a design solution space in advance. More recently, the concept of physics-informed neural network (PINN) has emerged as an alternative method for forward and inverse problems by satisfying governing partial differential equations (PDEs) across a range of engineering applications \cite{raissi2019physics, karniadakis2021physics, nair2024multiple,  kissas2020machine, lu2021physics}. For inverse design, while establishing a mathematical relationship that relates design parameters to target functionality is feasible in topology and shape optimization, this becomes significantly more difficult in placement optimization, especially in the case of multiphysics problems. 
Moreover, both data-driven DNNs and various forms of PINNs have been developed to function as surrogate forward solvers \cite{tao2019application, pestourie2020active, shukla2024deep, nair2024physics}. When combined with traditional optimization algorithms, these surrogates can accelerate the design optimization process. However, these surrogate-based inverse design methods still rely on traditional numerical solvers, thereby inheriting their limitations. Consequently, there is a need to explore dynamic placement optimization approaches that can effectively address challenges associated with nonlinear and multiphysics problems.

Reinforcement learning (RL) based machine learning models have been increasingly applied to solve a variety of design problems. While RL algorithms have been widely used in control theory and robotics \cite{brunke2022safe}, their application in engineering mechanics design has been limited \cite{dworschak2022reinforcement}.
In structural engineering, RL has been combined with graph embeddings for topology optimization, finding applications in the design of 1D bars and 2D structures \cite{hayashi2020reinforcement, sun2020generative}. In fluid dynamics, a design approach for active flow control introduced RL-based models \cite{rabault2019artificial}.  
Thereafter, the approach has been extended to various applications in fluid dynamics. In addition, Viquerat \etal \cite{viquerat2021direct} applied RL models for shape optimization using Bezier curves. More recently, Wu \etal \cite{wu2021design} presented a RL model for acoustic metamaterial design. However, a limited number of RL applications have addressed placement optimization problems subject to multiphysics constraints.

A well representative example of optimal design problems subject to multiphysics constraints is the design of microelectronic components. As demand for more efficient microchips grows, techniques for the rapid and accurate design and packaging of microelectronic components become more critical than ever. While traditional optimization techniques have been applied to heterogeneously integrated (HI) microelectronic systems under multiphysics constraints \cite{hadim2008multidisciplinary, zhao2016advanced, yu2022multi}, current electronic design automation (EDA) technologies face challenges in keeping up with the pace and scalability demands of modern microelectronic components \cite{yan2022towards}. Moreover, design optimization is often approached from a packaging perspective, focusing on minimizing geometry constraints rather than addressing multiphysics response constraints. While traditional solvers using heuristics and well-designed algorithms face limitations, machine learning (and RL in particular) has shown considerable promise in this area \cite{yan2022towards}.
Recent studies have employed RL for design optimization of microelectronic chips, especially in packaging, where RL models have been used to optimize power, performance, and footprint area (PPA) metrics \cite{mirhoseini2021graph}. These metrics are typically constrained by geometric factors such as placement density and wire routing congestion. Although these geometric factors can indirectly influence the multiphysics response of the design, RL models used in these studies are not directly constrained to satisfy specific multiphysics responses, which is crucial to achieve targeted physical responses. 

The primary goal of this study is to explore a class of RL frameworks to optimize microelectronic design under multiphysics constraints. We anticipate that, while we will specialize these constraints to be of thermoelastic type, the methodology will be general and applicable to other types of multiphysics problems. Two types of design problems typically occurring in microelectronics are investigated:
\begin{enumerate}
    \item \textit{Microchip interconnect design}: This problem consists of optimizing the bonded interconnect design of microchips under thermoelastic constraints. More specifically, the goal is to identify the optimal design parameters for interconnect configurations in an application-specific integrated circuit (ASIC) chip. Interconnects, which are essentially 'solder bumps' or 'solder joints', play a crucial role when connecting the microchip to the chip assembly. Their design is critical to ensure the reliable and efficient performance of the ASIC.
    The proposed RL model is tasked to determine the optimal interconnect configuration that minimizes temperature and mechanical stress levels. Moreover, this design problem is a combination of shape and placement optimization, as the RL model is responsible for identifying the optimal size and layout of interconnect arrays. To efficiently evaluate the physical responses, the RL model is integrated with a surrogate forward solver, which is a DNN pre-trained to map between the design parameters and their corresponding multiphysics responses.

    \item \textit{Heterogeneously integrated (HI) interposer design}: This problem consists of assembling a HI interposer to achieve an optimal design that satisfies thermoelastic constraints. The problem effectively presents a high-dimensional placement optimization task.
    While previous studies have addressed the packaging optimization of interposer models, they have primarily focused on enforcing geometric constraints \cite{mirhoseini2021graph}. To the best of our current knowledge, existing studies have not explored the optimization of HI interposer designs under thermoelastic response constraints.
    Most of the existing work in microelectronic packaging optimization for HI assemblies concentrated on both component placement and wire routing under geometry constraints to optimize power, performance, and footprint area (PPA). However, since component placement has a major impact on the physical response of the assembly, our study focuses primarily on optimizing the placement of components on the HI interposer to satisfy thermoelastic constraints. While the ASIC interconnect design problem can be addressed with a limited number of design variables in a discrete design space, the interposer design optimization is a high-dimensional problem that involves a continuous design space with a vast number of possible combinatorial actions and design parameters. Additionally, the placement of each component affects the position of the subsequent component, creating a dynamic design space where the optimization approach needs to keep track of the temporal dependencies between decisions and learn a strategy for optimal sequential decision-making. Moreover, the deep RL model also incorporates a surrogate forward solver in the form of a pre-trained DNN that maps the interposer design space to its corresponding multiphysics responses.
\end{enumerate}

The rest of the paper is organized as follows. First, \S\ref{ssec: Theory_TE} presents the theoretical background governing the thermoelastic response in solids. Next, \S\ref{ssec: fundamentals_RL} briefly introduces the fundamental concepts of reinforcement learning. Thereafter, \S\ref{ssec: ASIC_main} introduces the problem description and RL model implementation for ASIC interconnect design. Finally, \S\ref{ssec: Interposer_main} elaborates on the HI interposer design problem and highlights the deep RL model and its implementation. Both sections, \S\ref{ssec: ASIC_main} and \S\ref{ssec: Interposer_main}, are supported by corresponding numerical investigations to analyze the respective optimal design predictions (see \S\ref{ssec: ASIC_results} and \S\ref{ssec: Interposer_results}). 

\section{Theoretical preliminaries of thermoelasticity in solids}
\label{ssec: Theory_TE}

Consider a 3D solid domain represented by spatial coordinates $\mathbf{x}=(x, y, z) \in \Omega_0 \subset \mathbb{R}^3$. 
The heat transfer within the solid is assumed to occur according to a conduction mechanism and is described by the Fourier's law. For a stationary case, the thermal behavior of an isotropic solid is governed by the following steady-state heat conduction equation
\begin{equation}
    \label{eqn: GE_thermal}
    \nabla \cdot (-k \nabla T) = Q~~~~~~\mathbf{x} \in \Omega_0 
\end{equation}
where, $T$ is temperature, $k$ is the thermal conductivity, and $Q$ represents the heat per unit volume generated by the source.

Further, the equilibrium equation governing the static elastic response of an isotropic linear elastic solid under applied mechanical loads can be expressed as follows
\begin{equation}
    \label{eqn: GE_elastic}
    \nabla \cdot \mathbf{\sigma} + \mathbf{f} = 0~~~~~~\mathbf{x} \in \Omega_0 
\end{equation}
and
\begin{equation}
    \label{eqn: Const_eqm}
    \mathbf{\sigma} = \mathbf{C}:( \mathbf{\varepsilon} - \varepsilon_{th})
\end{equation}
where, $\sigma$ is the Cauchy stress tensor, $\mathbf{C}$ is the fourth order stiffness tensor, $\varepsilon = \frac{1}{2}[(\nabla \textbf{u})^T + \nabla \textbf{u}]$ is the strain tensor, $\varepsilon_{th} = \alpha_v (T-T_r)$ is the thermal strain tensor, and $\mathbf{f}$ are the body forces. In addition, $\textbf{u}$ is the displacement vector, $\alpha_v$ is the coefficient of thermal expansion, and $T_r$ is the reference temperature.

The thermoelastic response of solids is governed by Eqs.~\ref{eqn: GE_thermal}-\ref{eqn: GE_elastic} that describe the coupled thermal and mechanical behavior of the system. In addition to these governing equations, the problem is constrained by mechanical and thermal boundary conditions, which vary depending on the specific scenario. In this work, the boundary conditions are primarily enforced as follows
\begin{equation}
    \label{eqn: BC_thermoelasticity}
    \begin{split}
        \mathbf{u} &= 0~~~~~~\mathbf{x} \in \Gamma_c \\
        T &= T_0~~~~\mathbf{x} \in \Gamma_{i} \\
        -\textbf{n} \cdot \textbf{q} &=0~~~~~~\mathbf{x} \in \Gamma_{a}
    \end{split}
\end{equation}
where $\Gamma_c$ represents the fixed boundary, $\Gamma_i$ represents the isothermal boundary at a temperature $T_0$, $\Gamma_a$ represents a thermally insulated (adiabatic) boundary, $\mathbf{n}$ is the surface normal, and $\mathbf{q}$ is the thermal heat flux.

In the following section, we will briefly introduce the preliminaries of reinforcement learning. Thereafter, we will detail the specific RL models used for the microelectronic design problems under thermoelastic constraints.

\section{Fundamentals of reinforcement learning (RL)}
\label{ssec: fundamentals_RL}

Reinforcement learning (RL) is a type of machine learning model where an agent learns to make sequential decisions by interacting with an environment, receiving feedback in the form of rewards, and adjusting its actions accordingly. Unlike traditional neural network models that learn from static data, RL focuses on learning through exploration and interaction. Deep reinforcement learning (deep RL) enhances this approach by integrating deep learning techniques, allowing the model to handle complex, high-dimensional input spaces and enabling more sophisticated decision-making processes.

\begin{figure}[h!]
	\centering
	\includegraphics[width=1.0\linewidth]{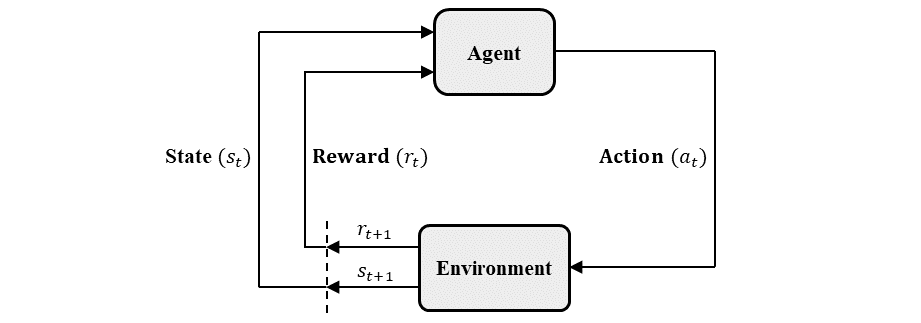}
	\caption{Schematic illustration of a basic reinforcement learning model highlighting the basic elements such as the agent, environment, action, state, and reward.}
	\label{fig: RL_fundamentals}
\end{figure}

Fundamentally, RL is modeled as a Markov decision process (MDP) \cite{sutton2018reinforcement} that consists of a set of \textit{states} ($\mathcal{S}$), a set of \textit{actions} ($\mathcal{A}$), the probability ($\mathcal{P}$) of transition from state $s_t$ to $s_{t+1}$ under action $a_t$ at time $t$, the set of \textit{rewards} ($\mathcal{R}$), and a discount factor $\gamma \in [0,1]$. The basic components of RL include the \textit{agent} and \textit{environment}, as shown in Fig.~\ref{fig: RL_fundamentals}. The environment represents the external system with which the agent interacts, while the agent is the entity responsible for making decisions based on the information it gathers. The agent interacts with the environment through a set of possible actions, denoted as $\mathcal{A}=[\mathcal{A}_1, \mathcal{A}_2,...]$. This action set encompasses all the potential actions the agent can take. The primary objective of reinforcement learning algorithms is to train the agent to select actions that maximize a predefined evaluation metric, known as the reward. The reward provides feedback to the agent, guiding it toward the optimal actions that achieve the desired outcome.

At an arbitrary time step $t$, the agent first observes the current state of the environment $s_t$ and the corresponding reward value $r_t$. Based on $s_t$ and $r_t$, the agent decided on an action $a_t$. This action is then executed within the environment, resulting in a new state $s_{t+1}$ and $r_{t+1}$. This process repeats iteratively, with the agent continuously refining its strategy to maximize the cumulative reward over time. The choice of strategy (or policy) depends on the problem at hand, with popular RL algorithms including Q-learning \cite{bellman1952theory} and policy gradient methods \cite{sutton1999policy}, and deep RL algorithms like deep Q-network (DQN) \cite{mnih2015human} and actor-critic method \cite{bahdanau2016actor}.

The learning process in RL involves balancing exploration (trying new actions to discover their effects) and exploitation (choosing actions that are known to yield high rewards). Through repeated interactions, the agent learns to associate specific actions with their corresponding rewards, enabling it to develop a policy that maps states to optimal actions.
In the following sections, we will elaborate on the development of RL models chosen to address the specific microelectronic design problems.

\section{ASIC interconnect design using RL}
\label{ssec: ASIC_main}

Application-specific integrated circuits (ASICs) are custom-designed electronic chips that integrate multiple circuits into one for a specific task or application. While an ASIC comprises both electrical and mechanical components, this study focuses on the mechanical design aspect. In particular, it addresses the thermoelastic response influenced by the thermal contributions of the electrical components. The ASIC mechanical design constraints for this study are established in accordance to application requirements for a general class of ASIC components. The primary objective is to design an optimal interconnect (or solder bump) configuration within the given design constraints of an ASIC, ensuring the proposed design satisfies given thermoelastic constraints. The interconnect configuration is a critical factor that determines the performance and reliability of an ASIC in microelectronic design. The goal is to develop a reliable interconnect design that avoids premature failures due to thermal loading based mechanical stresses. 

In the following sections, we first describe the geometry of the ASIC design. Then, we develop a finite element (FE) model to simulate its thermoelastic behavior. We introduce the RL framework to determine the optimal ASIC interconnect configuration. And finally, we evaluate the performance of the proposed RL based design optimization model by performing numerical simulations and analyzing the results.

\subsection{Geometric design for optimal thermoelastic response}

\begin{figure}[h!]
	\centering
	\includegraphics[width=1.0\linewidth]{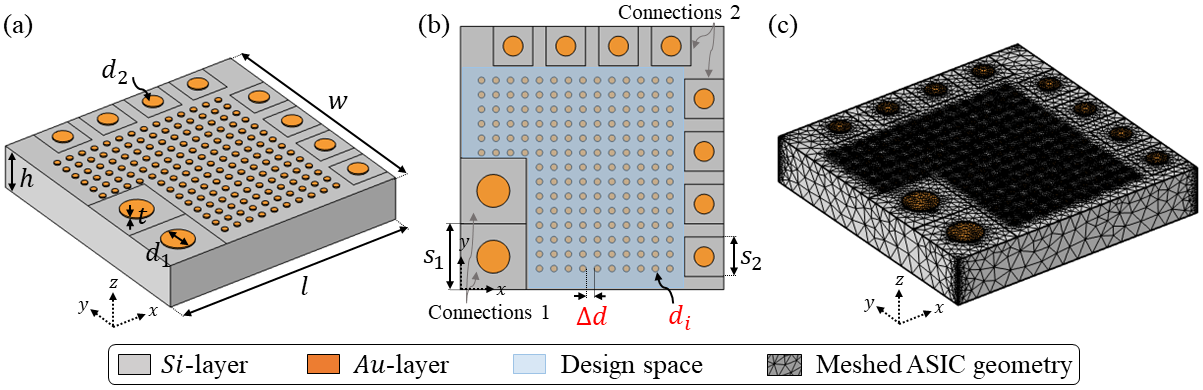}
	\caption{Schematic illustrating (a) the CAD model of 3D ASIC geometry, (b) the L-shaped design space (highlighted in light blue) for optimal interconnect configuration with design variables $d_i$ and $\Delta d$, and (c) the meshed 3D ASIC geometry for FE based multiphysics simulations. }
	\label{fig: ASIC_GeoMesh}
\end{figure}

The design problem treated as a benchmark for this study is illustrated in Fig.~\ref{fig: ASIC_GeoMesh}. Fig.~\ref{fig: ASIC_GeoMesh}(a) highlights the geometry and material distribution of the ASIC, the design space and variables that are subject to optimization are highlighted in Fig.~\ref{fig: ASIC_GeoMesh}(b). In this work, the silicon chip is denoted as the $Si$-layer. The $Au$-layer, on the other hand, represents the remaining geometry made of gold material, including the interconnects and ten additional electrical connections (enclosed within square boxes in Fig.~\ref{fig: ASIC_GeoMesh}(b)). 
According to manufacturing constraints, the interconnect configuration must consist of interconnects of uniform size arranged in a periodic configuration within a L-shaped design space as highlighted in Fig.~\ref{fig: ASIC_GeoMesh}(b), that is all the interconnects in a given configuration must have the same optimal diameter with constant spacing. Therefore, this design problem aims to determine the optimal interconnect diameter ($d_{i}$) and edge-to-edge spacing ($\Delta d$) such that the resulting ASIC configuration satisfies temperature and stress constraints. Specifically, the optimal configuration should have temperature and stress levels below the maximum allowable values of $T_0$ and $\sigma_0$, respectively. Note that, for the numerical results presented in the following, these constraint values are $T_0=358.15~K$ ($85~^{\circ}C$), which is the maximum allowable temperature on the $Si$ surface, and $\sigma_0=200~MPa$, which is the yield stress limit of the $Au$ interconnect. In addition, the design variables must adhere to the following ranges: $d_i \in [10, 50]~\mu m$ and $\Delta d \in [20, 100]~\mu m$. The fixed geometry and material parameters used in this problem are tabulated in Table~\ref{table: Table_ASIC}. 

\begin{table}[ht]
    \centering
    \begin{tabular}{|c|c|c|c|}
        \cline{2-4}  
        \multicolumn{1}{c}{} & \multicolumn{1}{|c|}{\textbf{Parameter}} & \multicolumn{1}{c|}{\textbf{$Si$-layer}} & \multicolumn{1}{c|}{\textbf{$Au$-layer}} \\ 
        \hline
		 & Length - $l$ ($\mu m$) & 2000 & - \\
         & Width - $w$ ($\mu m$) & 2000 & - \\
         & Height - $h$ ($\mu m$) & 350 & - \\
         & Enclosure side length for connections 1 - $s_1$ ($\mu m$) & 500 & - \\
         \textbf{Geometry} & Enclosure side length for connections 2 - $s_2$ ($\mu m$) & 300 & - \\
         & Diameter of connections 1 - $d_1$ ($\mu m$) & - & 250 \\
         & Diameter of connections 2 - $d_2$ ($\mu m$) & - & 150 \\
         & Thickness of connections - $t$ ($\mu m$) & - & 10 \\
         & and interconnects &  &  \\
        \hline
         & Density - $\rho$ ($\frac{kg}{m^3}$) & 2329 & 19300 \\
         & Young's modulus - $E$ ($GPa$) & 170 & 70 \\
         \textbf{Material} & Poisson's ratio - $\nu$ & 0.28 & 0.44 \\
         & Thermal conductivity - $\kappa$ ($\frac{W}{mK}$) & 131 & 317 \\
         & Coeff. of thermal expansion - $\alpha_v$ ($\frac{1}{K}$) & 2.6e-6 & 14.2e-6 \\
        \hline
    \end{tabular}
    \caption{Summary of fixed geometry and material parameters used to develop the FE model to simulate the thermoelastic response of the ASIC.}
    \label{table: Table_ASIC}
\end{table}

\subsection{Thermoelastic FE model of the ASIC}
\label{ssec: FEModel_ASIC}

An essential aspect of the design optimization under physical constraints is the implementation of a forward solver to evaluate the physical response of the system corresponding to each iteration produced by the optimization process. To analyze the thermoelastic response of the ASIC, we develop and simulate a 3D ASIC model using the finite element (FE) software COMSOL Multiphysics$\textsuperscript{\textregistered}$ as shown in Fig.~\ref{fig: ASIC_GeoMesh}(c). The geometry and material parameters provided in Table~\ref{table: Table_ASIC} are used to build the FE model. 

The forward model is set up to evaluate the stationary thermoelastic response of the ASIC by solving Eqs.\ref{eqn: GE_thermal}-\ref{eqn: BC_thermoelasticity} introduced in \S\ref{ssec: Theory_TE}. Adiabatic thermal boundary conditions are applied to the $Si$-layer, while isothermal boundary conditions (at $T=343.15~K$) are enforced on the upper surfaces of all the cylindrical structures in the $Au$-layer. Additionally, the $Si$-layer serves as a thermal heat source with a power dissipation value of $P=415~mW$. From a mechanical point of view, the ASIC geometry is constrained via fixed boundary conditions on the upper surfaces of all the cylindrical structures in the $Au$-layer to simulate the attachment of the ASIC chip to a microelectronics assembly. Further, the developed model is meshed with tetrahedral elements (see Fig.~\ref{fig: ASIC_GeoMesh}(c)) with a minimum element size of $\Delta h =5~\mu m$. The FE model is simulated for a combination of $d_i$ and $\Delta d$ and the corresponding maximum temperature ($T_m$) on the $Si$-layer and maximum stress ($\sigma_m$) on the $Au$-layer are recorded for each simulation. Here, $\sigma_m$ refers to the maximum \textit{von Mises stress} in the design, and for the sake of brevity, it will be simply referred to as \textit{stress} in the following.

\subsection{Modeling Approach}
\label{ssec: ASIC_RLModel}

\begin{figure}[h!]
	\centering
	\includegraphics[width=1.0\linewidth]{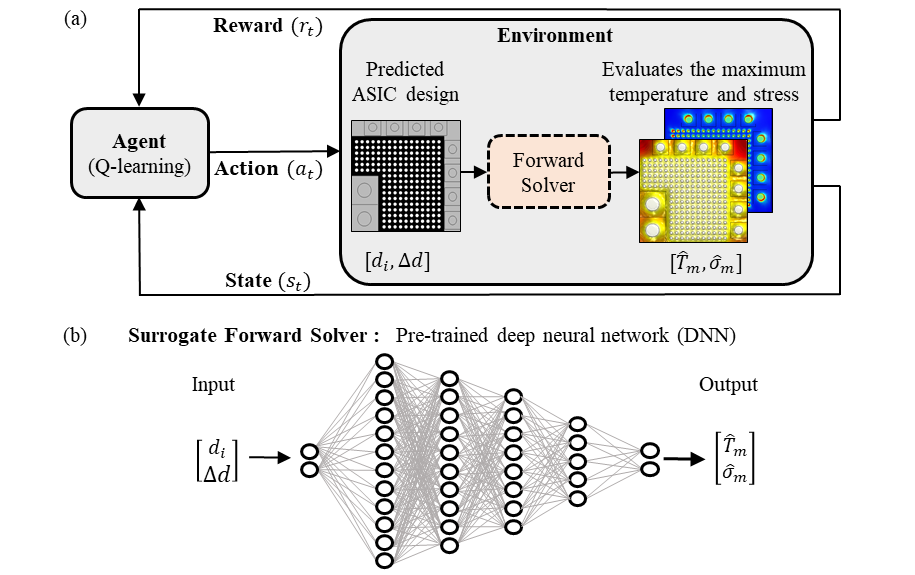}
	\caption{Schematic illustrating (a) the RL algorithm implementation, and (b) the fully connected deep neural network (DNN) model trained to be the surrogate forward model. }
	\label{fig: RL_ASIC}
\end{figure}

We propose a reinforcement learning (RL) framework for the design of optimal interconnect configuration in ASIC that is schematically illustrated in Fig.~\ref{fig: RL_ASIC}(a). The proposed approach uses RL to estimate the design parameters capable of delivering a targeted physical behavior (e.g., satisfying temperature and stress limits). During the design process, the RL agent selects an optimal action ($a_t$) for the current state ($s_t$) based on a strategy, also called a policy ($\pi$), to choose the design parameters and interact with the environment. In the specific case of this study, the RL environment is the forward thermoelastic model with which the agent interacts to evaluate the temperature and stress levels. Further, a reward ($r_t$) is assigned to the agent based on the chosen $a_t$. Starting with an initial guess of design parameters, the RL algorithm iteratively searches for the optimal parameters based on the $r_t$ estimated on the basis of the physical responses calculated by a forward solver. The agent continues exploring the design space until the target conditions are satisfied. In this study, the design process ends once the thermoelastic response of the ASIC satisfies a target behavior.

As discussed above, the RL framework has two integral components: the forward solver and the RL algorithm. Therefore, the complete RL framework will be presented in two separate sections. The first section will introduce the forward solver embedded within the RL model, which is crucial for calculating $r_t$ for optimal design parameters. The second part will elaborate on the details of the RL algorithm used and the optimization process followed.

\subsubsection{DNN based surrogate model employed as forward solver}
\label{ssec: ASIC_SurrogateModel}

\begin{figure}[h!]
	\centering
	\includegraphics[width=1.0\linewidth]{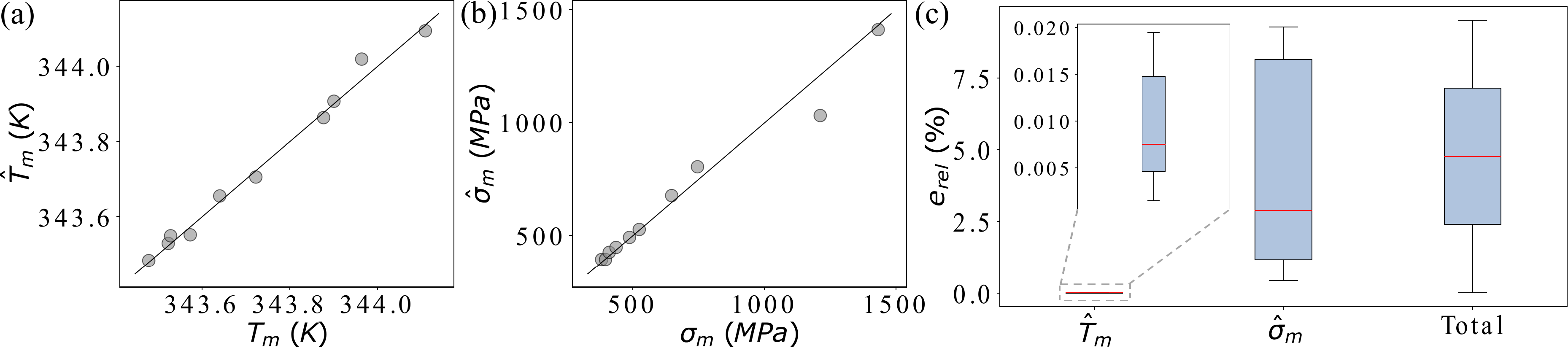}
	\caption{Plots showing the comparison between the predictions of the trained surrogate model with the ground truth based on test dataset for (a) temperature ($T_m$) and (b) stress ($\sigma_m$) fields. The closer the predictions are to the dashed line the better the prediction accuracy of the surrogate model. (c) Box plots highlighting the relative error $e_{rel}$ (in $\%$) distribution across the test dataset for $T_m$, $\sigma_m$, and $T_m -\sigma_m$ together (total). Note that the red line in each box plot represents the median $e_{rel}$. }
	\label{fig: ASICSurr_data}
\end{figure}

While the RL model commonly employs traditional forward solvers (e.g. FEM), the integration of a traditional solver significantly increases the computational time of the whole design optimization process. To address this issue, we develop a surrogate forward solver that is based on a deep neural network (DNN) architecture. Specifically, we design a feed-forward DNN that is trained to accurately predict $\hat{T}_m$ and $\hat{\sigma}_m$ based on the design parameters ($d_i$ and $\Delta d$) provided as input to the model, as shown in Fig.~\ref{fig: RL_ASIC}(b). Note that the quantities marked with $\hat{(.)}$ represent the surrogate model predictions. Once trained, the DNN can be leveraged to act as a surrogate forward solver within the RL framework (see Fig.~\ref{fig: RL_ASIC}).

The surrogate model is a fully connected neural network architecture that consists of four hidden layers, each with 25 neurons, followed by $ReLU$ activation functions in each layer. The training data for the surrogate DNN model is generated using the FE model presented in \S\ref{ssec: FEModel_ASIC}. The FE model is used to generate the response of 500 different interconnect configurations under varying combinations of $d_i$ and $\Delta d$, and the corresponding $T_m$ and $\sigma_m$ values are recorded for each simulation. Using this labeled input-output dataset, the network is trained to find the optimal network parameters $\theta$ by minimizing the mean square error (MSE) loss function as follows
\begin{equation}
\begin{split}
    \mathcal{L}(\theta) =&~ \omega_1 \mathcal{L}_T(\theta) + \omega_2 \mathcal{L}_{\sigma}(\theta) \\
    =&~ \frac{\omega_1}{N} \sum^N_{i=1} \Big( \hat{T}_m(\theta) - T_m \Big)^2 + \frac{\omega_2}{N} \sum^N_{i=1} \Big( \hat{\sigma}_m(\theta) - \sigma_m \Big)^2
\end{split}
\end{equation}
where $\mathcal{L}_T$ is the MSE loss for temperature and $\mathcal{L}_{\sigma}$ is the MSE loss for stress. In addition, the network is trained with $N$ training data samples with $\omega_1=1$ and $\omega_2 =1$ as the weighting factors for $\mathcal{L}_T$ and $\mathcal{L}_{\sigma}$, respectively. Although temperature and stress are defined on different units and numerical scales, the same weighting factors were used during training because both physical properties are normalized. This normalization ensures that despite their different magnitudes, temperature and stress are treated equivalently during the network training.

The total dataset of 500 samples is split into datasets of $N=400$ samples for training, $N_{val}=90$ samples for validation, and $N_{test}=10$ samples for testing. The DNN is trained using Adam optimizer with a learning rate of 1e-3, batch size of 32 samples for 500 epochs, and implemented in Python 3.8 using Pytorch API on NVIDIA A100 Tensor Core GPU with 80GB memory. 

In addition, the performance of the trained surrogate model is evaluated qualitatively by plotting and comparing the model prediction with the ground truth for both the temperature and stress fields in Fig.~\ref{fig: ASICSurr_data}(a) and (b), respectively. The closer the predictions are to the dashed straight lines, the higher the chances that the predictions and ground truth belong to the same distribution. This comparison indicates low variability and high accuracy in the network prediction. This is corroborated quantitatively by the box plot in Fig.~\ref{fig: ASICSurr_data}(c). The box plot highlights the variation of the relative $\%$ error ($e_{rel}$) in the prediction of temperature, stress, and temperature-stress (considered together -- \enquote{Total}). Moreover, calculating the mean prediction error across the test dataset, evaluates $\bar{e}_{rel} |_{T_m} = 0.01 \%$, $\bar{e}_{rel} |_{\sigma_m} = 9.51 \%$, and $\bar{e}_{rel} |_{T_m-\sigma_m} = 4.76 \%$. While the overall relative error $\bar{e}_{rel} |_{T_m-\sigma_m}$ across the test dataset indicates high prediction accuracy, the relative errors in temperature and stress predictions vary significantly. This variation can be attributed to the differences in ranges of temperature and stress in the training data. While the temperature varies only about $1~K$ for the proposed ASIC designs, the variation of the stress for the same models is on the order of $1000~MPa$. Despite data normalization during training, smaller variations in temperature lead to higher prediction accuracy as indicated by smaller $\bar{e}_{rel} |_{T_m}$, whereas larger variation in stress results is slightly higher $\bar{e}_{rel} |_{\sigma_m}$. However, the low overall prediction error $\bar{e}_{rel} |_{T_m-\sigma_m}$ highlights the good prediction accuracy of the proposed surrogate model.

\subsubsection{Q-learning algorithm}

The RL framework (see Fig.~\ref{fig: RL_ASIC}(a)) is developed using a Q-learning algorithm. The model attempts to find the state ($s_t$) by selecting optimal actions ($a_t$) to increase or decrease $d_i$ and $\Delta d$ as follows
\begin{equation}
  \label{eqn: state_asic}
  S(s_t) = (d_i, \Delta d)~~s.t.~
    \begin{cases}
      d_i \in [10, 50]\mu m\\
      \Delta d \in [20, 100]\mu m\\
    \end{cases}       
\end{equation}
and
\begin{equation}
    \label{eqn: action_asic}
    A(a_t) = \big[+\delta d_i, -\delta d_i, +2 \delta \Delta d, -2 \delta \Delta d \big]
\end{equation}
where $\delta(.)=\frac{40}{99}~\mu m$ is the spatial search discretization chosen for this study based on the manufacturing limits of the ASIC layouts. The state-action pairs are represented $S(s_t)-A(a_t)$ with four possible scenarios in the action space $A$ at each state $s_t$ in the state space $S$. 

The overall goal to determine the optimal $a_t$ at the current $s_t$ is evaluated using the Q-learning algorithm. The algorithm updates the $Q$ value of each $s_t-a_t$ pair using the Bellman equation \cite{bellman1952theory} as follows 
\begin{equation}
    \label{eqn: Q_algo}
    Q(s_t, a_t) \leftarrow Q(s_t, a_t) + \alpha_r \Big( r_t + \gamma~ \max_{a'_t \in A} Q(s_{t+1}, a'_t) - Q(s_t, a_t) \Big)
\end{equation}
where $Q(s_t, a_t)$ is the updated $Q$ value for picking action $a_t$ at state $s_t$, $\alpha_r=0.9$ is the learning rate, $r_t$ is the immediate reward for choosing $a_t$, $\gamma=0.99$ is the discount factor for expected future rewards, and $\max_{a'_t \in A} Q(s_{t+1}, a'_t)$ is the maximum $Q$ value among all actions $a'_t\in A$ for the next state $s_{t+1}$. In this study, $r_t$ is defined as a function of $\hat{T}_m$ and $\hat{\sigma}_m$ as follows
\begin{equation}
    \label{eqn: r_asic}
    r_t = \alpha \frac{T_0 - \hat{T}_m}{T_n} + (1-\alpha) \frac{ \sigma_0 - \hat{\sigma}_m}{\sigma_n}
\end{equation}
where $T_0$ and $\sigma_0$ represent the limiting temperature and yield stress for the ASIC design, $T_n=1e-6K$ and $\sigma_n=1e4~Pa$ represent the normalization temperature and stress, and $\alpha$ and $1-\alpha$ are the weight factors for temperature and stress, respectively. The normalization factors are chosen by trial and error such that both the temperature and stress related terms in $r_t$ are of of comparable order. In addition, $\alpha \in [0,1]$ such that $\alpha=0$ represents a temperature-independent reward and $\alpha=1$ represents a stress-independent reward function. Note that $\hat{T}_m$ and $\hat{\sigma}_m$ for each $s_t-a_t$ pair is evaluated using the trained surrogate model. While the $Q$ values are initialized as zero for all $s_t-a_t$ pairs, the model iteratively learns and updates the $Q$ values as per Eq.~\ref{eqn: Q_algo}. 

The RL model is trained by maximizing the $r_t$ to determine the optimal $d_i$ and $\Delta d$ values that meet the limiting $T_0$ and $\sigma_0$ constraints for the ASIC. At its core, the Q-learning algorithm is designed to maximize the $r_t$ by learning to select the actions with the maximum reward path. However, this reward \textit{greedy} approach can cause the algorithm to focus only on the immediate high-reward actions, potentially missing out on better solution paths with higher cumulative rewards. As a result, the algorithm may converge to a local optimum, rather than finding the global optimum in the design space.

\begin{algorithm}[H]
    \caption{The pseudocode for the proposed $\epsilon$-greedy Q-learning based RL algorithm}\label{alg: eps_Qlearning}
    Pick random initial state $s^0_t$\;
    Initialize $Q(s_t, a_t)=0$\;
    $\epsilon=0.1$, $\alpha_r=0.9$, $\gamma=0.99$\; 
    \For{episode = $1$ to $N$}{
        $s_t \gets s^0_t$\;
        \While{$s_t$ is not terminal state}{
            Generate a random number $r\in[0,1]$\;
            \eIf{$r<\epsilon$}{
                Select an exploratory action $a_t$\;}
            {Select a greedy action $a_t$\;
            }    
            Determine $s_{t+1}$ based on $s_t$ and $a_t$\;
            Evaluate $r_t$ using Eq.~\ref{eqn: r_asic}\;
            Update $Q(s_t, a_t)$ using Eq.~\ref{eqn: Q_algo}\;
            $s_t \gets s_{t+1}$\;
            }
        Find the optimal state $s^*_t$ by locating the maximum $Q$-score\;
        $s^0_t \gets s^*_t$;
    }  
\end{algorithm}

We address this challenge by employing the $\epsilon$-greedy Q-learning algorithm \cite{wunder2010classes} as described in Algorithm~\ref{alg: eps_Qlearning}. While the implementation remains similar to the basic Q-learning algorithm, the $\epsilon$-greedy learning process introduces a balance between exploitation (choosing the high reward action) and exploration (trying new actions). In this approach, the agent chooses a policy $\pi$ to either select the greedy $a_t$ (exploitation) that maximizes $Q$-score at the current state $s_t$ or pick an exploratory action that may lead to unexplored $s_t$. In each iteration, a random number $r\in [0,1]$ is generated. An exploratory action is selected if $r<\epsilon$, otherwise a greedy action is chosen.
In our case, we set $\epsilon=0.1$ which implies a $10\%$ probability of choosing a random exploratory action, regardless of the current $Q$ value. This exploration-exploitation strategy enables the agent to avoid getting trapped in local optima and increases the likelihood of discovering actions that can lead to a higher overall reward.

\subsection{Numerical experiments and discussion}
\label{ssec: ASIC_results}

In this section, we evaluate the performance of the proposed RL framework, primarily focusing on its ability to identify the optimal design parameters for the ASIC under thermoelastic constraints. We begin by detailing the optimization process, followed by an analysis of the optimal designs predicted by the RL models. Finally, we validate these results using FE simulations.

\subsubsection{Optimization process}

\begin{figure}[h!]
	\centering
	\includegraphics[width=1.0\linewidth]{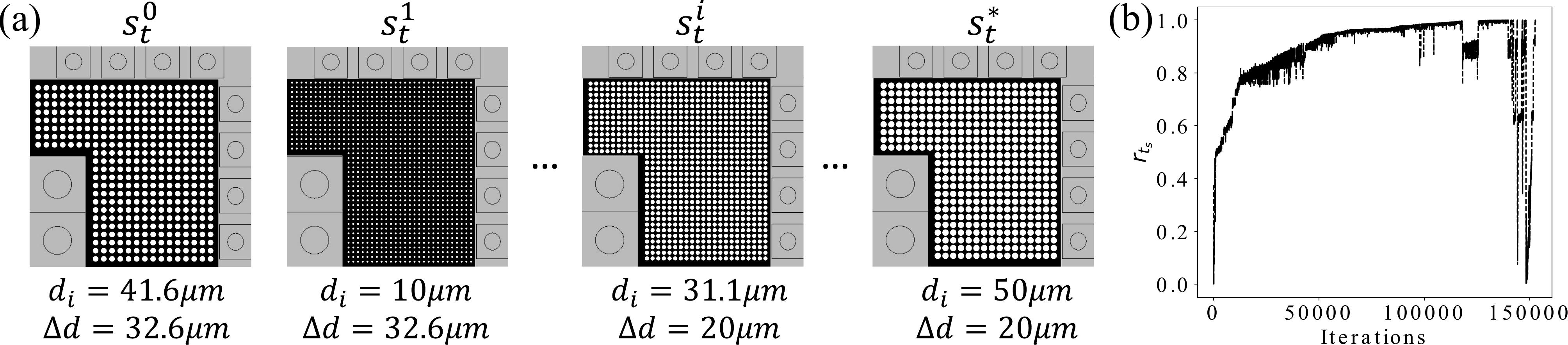}
	\caption{(a) The evolution of the ASIC interconnect configuration during the optimization process. The optimization process starts at an initial state $s^0_t$, and attempts to find the optimal interconnect design parameters ($d_i$ and $\Delta d$) by the terminal state $s^*_t$. (b) The plot describes the variation of the reward function with iterations. Note that a scaled reward function $r_{t_s}$ is plotted such that $r_{t_s}: r_t \rightarrow [0,1]$.}
	\label{fig: RL_iters}
\end{figure}

As discussed earlier, the RL framework identifies the optimal design parameters by maximizing $r_t$. A higher $r_t$ value enables the optimization model to find designs with $T_m$ and $\sigma_m$ close to the limiting values of $T_0$ and $\sigma_0$. The RL framework iteratively learns to converge to optimal $d_i$ and $\Delta d$ values corresponding to maximum $r_t$. This iterative evolution process is schematically represented in Fig.~\ref{fig: RL_iters}(a). The model begins from an initial state $s^0_t$ and iteratively passes through various states within the design space over multiple training episodes, eventually converging to an optimal state $s^*_t$ associated with maximum $r_t$.
Additionally, Fig.~\ref{fig: RL_iters}(b) illustrates the variation in reward function with iterations, thereby quantifying the RL based learning process. For ease of representation, the final reward $r_t$ is mapped to a scaled reward $r_{t_s}$ such that $r_{t_s}: r_t \rightarrow [0,1]$. Initially, the RL model starts from design parameters that yield a small $r_t$. However, as the model iterates over different training episodes, it learns to identify a path that maximizes $r_t$. 
As observed throughout the learning process and particularly towards the final iterations (see Fig.~\ref{fig: RL_iters}(b)), the $\epsilon$-greedy Q-learning RL model explores the design space from different initial points but consistently finds the optimal design parameters. The model is considered to have converged to an optimal solution when it observes the same maximum reward across multiple iterations. 

\subsubsection{Optimal ASIC interconnect designs and validation}

\begin{figure}[h!]
	\centering
	\includegraphics[width=1.0\linewidth]{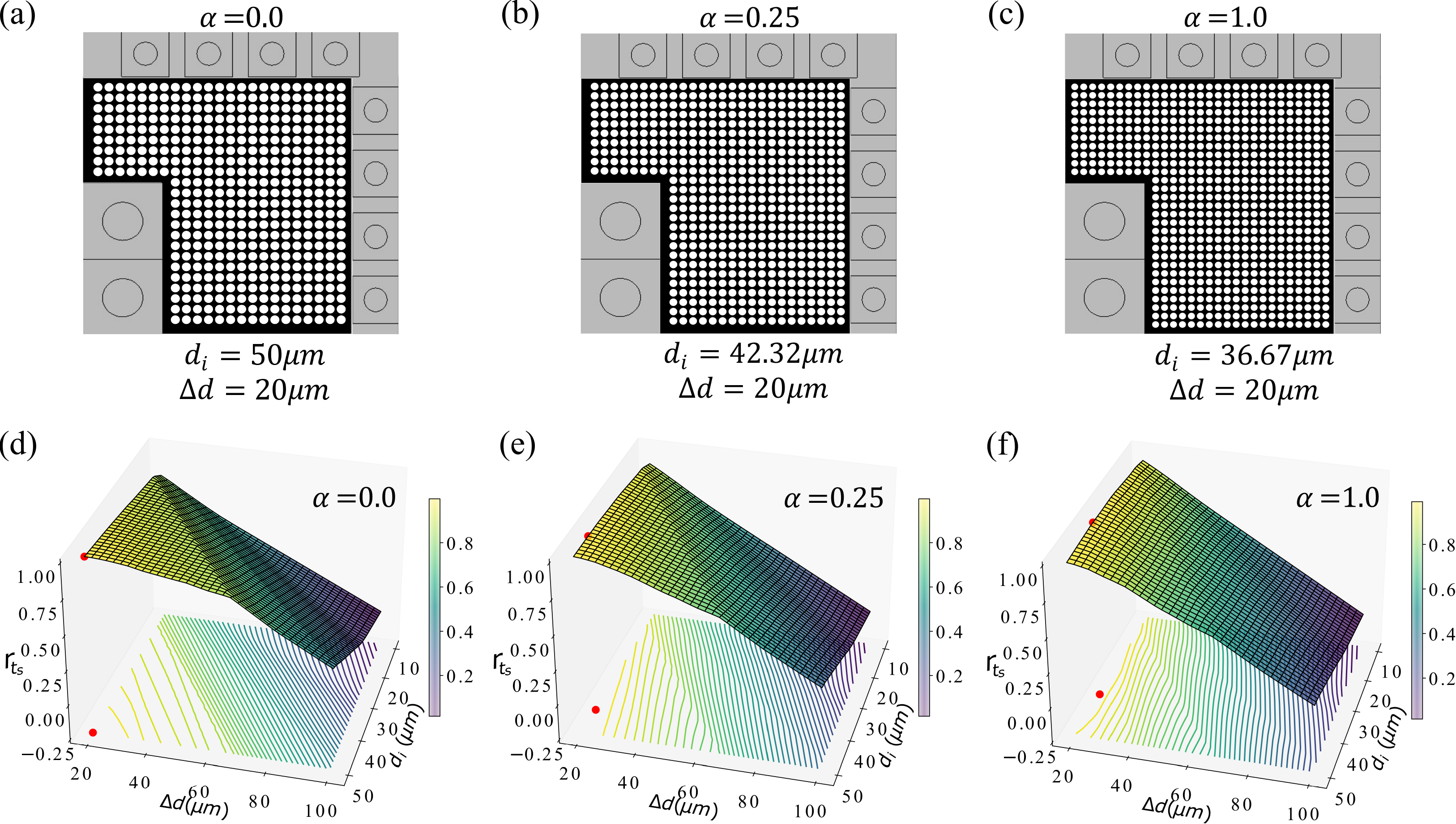}
	\caption{Schematic demonstrates the optimal ASIC designs for (a) $\alpha=0.0$, (b) $\alpha=0.25$, and (c) $\alpha=1.0$. In addition, the 3D contour plots (with corresponding 2D projections) represent the variation of $r_{t_s}$ with $d_i$ and $\Delta d$ for (d) $\alpha=0.0$, (e) $\alpha=0.25$, and (f) $\alpha=1.0$. The red circle markers on the contour plots indicate the respective maximum $r_{t_s}$ points, which coincide with the optimal design parameters predicted by the RL model in (a)-(c). 
    Note that a scaled reward function $r_{t_s}$ is plotted such that $r_{t_s}: r_t \rightarrow [0,1]$.
    }
	\label{fig: designSol_ASIC}
\end{figure}

\begin{figure}[h!]
	\centering
	\includegraphics[width=1.0\linewidth]{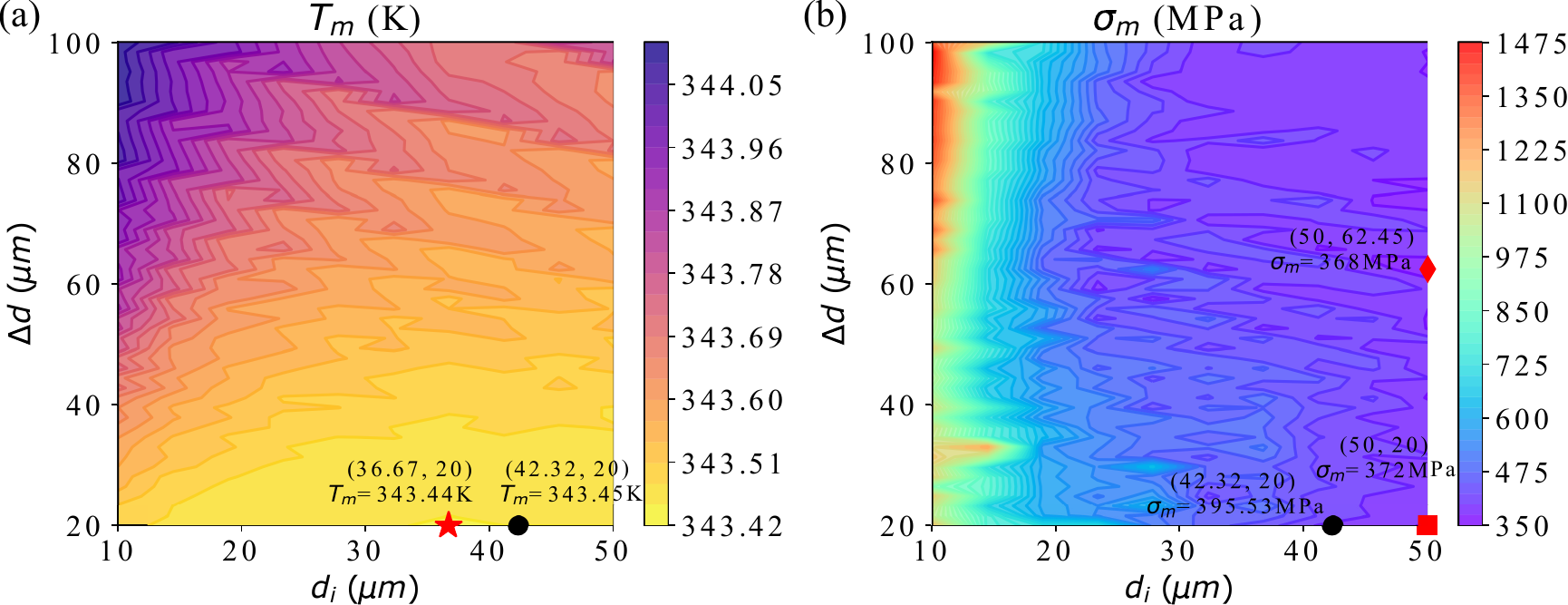}
	\caption{Contour plots showing the variation of (a) the maximum temperature $T_m$ (in $K$) and (b) the maximum stress $\sigma_m$ (in $MPa$) for different ASIC interconnect designs, obtained using FE simulations. Both the contour plots highlight the variation of the true physical quantities with respect to the $d_i$ and $\Delta d$. In addition, the markers indicate the optimal points under different optimization constraints. The red star marker indicates the true optimal parameter values under a thermal constraint alone, the red diamond and square markers correspond to the true and predicted optimal parameter values under a stress constraint alone, and the black circular markers indicate the optimal predicted parameter values under a combined thermoelastic constraint.}
	\label{fig: FEMsol_ASIC}
\end{figure}

In this section, the performance of the proposed model in terms of determining the optimal design parameters ($d_i$ and $\Delta d$) is evaluated. Since this study involves a multi-objective reward function (Eq.~\ref{eqn: r_asic}) with both temperature and stress constraints, the weighting factor ($\alpha$) plays a critical role in determining the optimal design. In other words, the optimal design parameters can vary depending on the selected value of $\alpha$ in Eq.~\ref{eqn: r_asic}. Therefore, we analyze the performance of the optimization model for different values of $\alpha \in [0,1]$ as illustrated in Fig.~\ref{fig: designSol_ASIC}(a)-(c).
Additionally, we examine 3D contour plots showing the variation of the scaled reward ($r_{t_s}$) with the design parameters, as seen in Fig.~\ref{fig: designSol_ASIC}(d)-(f), to ensure that the optimal $d_i$ and $\Delta d$ values correspond to the maximum $r_{t_s}$ at different $\alpha$ values. Finally, the RL predictions are validated using 2D contour plots (obtained from FE analyses) depicting the true variation of $T_m$ and $\sigma_m$ with the design parameters, as shown in Fig.~\ref{fig: FEMsol_ASIC}(a) and Fig.~\ref{fig: FEMsol_ASIC}(b). 

We begin by analyzing the performance of the RL model in the extreme cases of $\alpha=0$ and $\alpha=1$. First, we consider $\alpha=0$ in Eq.~\ref{eqn: r_asic}, which allows focusing on the ability of the RL model to predict the optimal design parameters under elastic constraints only. In this scenario, the RL model is trained to identify the optimal design parameters that minimize mechanical stress under thermal loading, independent of the corresponding temperature rise in the ASIC.
The RL model predicts an optimal state $s^*_t$ with design parameters $d_i=50~\mu m$ and $\Delta d=20~\mu m$ for $\alpha=0$. A comparison with the true $\sigma_m$ contour plot in Fig.~\ref{fig: FEMsol_ASIC}(b) highlights the accuracy of this prediction. The true optimal design parameters are $d_i=50~\mu m$ and $\Delta d=62.45~\mu m$ (indicated by the red diamond marker in Fig.~\ref{fig: FEMsol_ASIC}(b)). However, the $\Delta d$ prediction of the RL model (indicated by the red square marker in Fig.~\ref{fig: FEMsol_ASIC}(b)) is different from the true optimal value.
Although this might stand out as a significant error in the identification of the design parameters, it is important to highlight that the reward function in this case is stress-dependent. Therefore, comparing the error in stress values between the true and predicted design parameters provides a more accurate assessment of the model performance. 
To this end, a careful investigation of the true $\sigma_m$ contour plot in Fig.~\ref{fig: FEMsol_ASIC}(b) reveals that the resulting relative error in stress values between the predicted and the true optimal design parameters is only $e_{rel}=1.08 \%$. This discrepancy can be attributed to the prediction error of the surrogate forward solver, where small inaccuracies in $\hat{T}_m$ and $\hat{\sigma}_m$ predictions could result in an overall prediction error in the RL framework. Enhancing the accuracy of the forward solver could further improve the overall accuracy of the RL model. However, note that the current relative prediction error ($e_{rel} < 1.1 \%$) between stresses in true and predicted design parameters at $\alpha=0$ is still insignificant.
While the RL model effectively identifies a close global minimum of the $\sigma_m$ by maximizing the $r_t$, finding this global minimum using traditional optimization methods can be significantly more challenging. This difficulty arises from the fact that the true optimal design parameters lie within a region of the design space characterized by non-convex stress field as shown in Fig.~\ref{fig: FEMsol_ASIC}(b).

Second, we evaluate the ability of the RL model to predict optimal design parameters for $\alpha=1$ in Eq.~\ref{eqn: r_asic}. This scenario focuses on the capability of the RL model to determine the optimal design under thermal constraints only. In this case, the RL model is trained to solve for the optimal design parameters that minimize the temperature rise under thermal loading, independent of the mechanical stress produced in the ASIC. The RL model predicts an optimal state $s^*_t$ with design parameters $d_i=36.67~\mu m$ and $\Delta d=20~\mu m$ for $\alpha=1.0$, as shown in Fig.~\ref{fig: designSol_ASIC}(c) and (f). A direct comparison with the true $T_m$ contour plot in Fig.~\ref{fig: FEMsol_ASIC}(a) confirms the high prediction accuracy of the model. 
The prediction accurately matches the true optimal design parameters, indicated by the red star marker in Fig.~\ref{fig: FEMsol_ASIC}(a). As a result, the RL model prediction will lead to the minimum temperature rise in the ASIC, with $T_m=343.44~K$. Notably, the prediction accuracy for $\alpha=1$ is higher than that for $\alpha=0$ due to the relatively smoother $T_m$ (Fig.~\ref{fig: FEMsol_ASIC}(a)) field compared to the $\sigma_m$ (Fig.~\ref{fig: FEMsol_ASIC}(b)) field.

Finally, we assess the ability of the RL model to predict optimal design parameters for the multi-objective $r_t$ that incorporates both temperature and stress constraints. For instance, Fig.~\ref{fig: designSol_ASIC}(b) and (e) demonstrate the ability of the model to identify the optimal parameters when $\alpha=0.25$, where higher $\alpha$ gives higher emphasis to the temperature dependence. In this case, the RL model predicts optimal design parameters of $d_i=43.32~\mu m$ and $\Delta d=20~\mu m$, as shown by the black circular markers in the true $T_m$ and $\sigma_m$ contour plots in Fig.~\ref{fig: FEMsol_ASIC}. When the FE model is simulated with the predicted design parameters, we evaluate $T_m=343.45~K$ and $\sigma_m = 395.53~MPa$ for the optimal model. 
As $\alpha$ varies between 0 and 1, the optimal design parameters shift between the two extreme cases of purely temperature-dependent (red star marker in Fig.~\ref{fig: FEMsol_ASIC}(a)) and purely stress-dependent (red square marker in Fig.~\ref{fig: FEMsol_ASIC}(b)). These results underscore the ability of the RL model to navigate a complex design space and find optimal design parameters.
The high prediction accuracy demonstrates the effectiveness of the proposed RL framework in solving the complex ASIC design problem under multiphysics constraints.


\section{Interposer placement optimization using deep RL}
\label{ssec: Interposer_main}

Another key element in the design of microelectronic components is the heterogeneously integrated (HI) interposer assembly. The HI interposer integrates separately manufactured components into a high-level assembly, enhancing both functionality and operational characteristics. Heterogeneous integration allows for the combination of different types of components within a single package. Additionally, by incorporating components with varying power requirements, the HI interposer can significantly improve overall energy efficiency.

HI interposers are essential for enabling advanced applications that demand high levels of integration and performance, achieved by combining multiple chip types in a single package. This can be viewed as an electronic packaging problem. Traditional electronic packaging problems primarily focus on minimizing wire routing for optimal power efficiency without directly considering physical constraints. However, our focus is on the optimal placement of components within the HI interposer such that the design assembly adheres to specific physical response constraints. In particular, the optimal HI interposer design should satisfy multiphysics constraints to ensure that the interposer experiences minimal temperature and stress increases while requiring a minimum footprint. Note that the terms HI interposer and interposer will be used interchangeably in the following.

We first describe the geometry of the HI interposer design, then we discuss a finite element (FE) model to simulate the thermoelastic behavior of the interposer design. With the FE model available, we shift our focus on the RL framework used to determine the optimal interposer assembly configuration. Finally, we evaluate the performance of the proposed RL based design optimization model by applying it to numerical test cases.

\subsection{Geometric design for optimal thermoelastic response}

\begin{figure}[h!]
	\centering
	\includegraphics[width=1.0\linewidth]{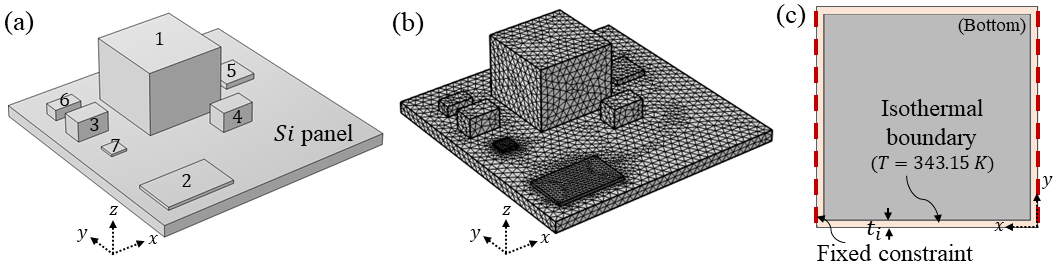}
	\caption{Schematic illustrating (a) the CAD model of a 3D HI interposer geometry with seven components (marked 1-7) placed on the $Si$ panel, (b) the meshed interposer geometry used for the FE based thermoelastic simulations, and (c) the isothermal boundary condition on the square annulus (highlighted) and fixed constraint on the left and right edges (red dashed lines) of the bottom surface of the interposer model.}
	\label{fig: Inter_Geo}
\end{figure}

In this study, we consider mechanical design criteria of a generic and standard HI interposer model. We aim to develop an optimization approach to determine the optimal design of an interposer assembly comprising $N_c=7$ components, as shown in Fig.~\ref{fig: Inter_Geo}(a). Specifically, this study focuses on an interposer that integrates the following components: one feedback controller chip, ASIC, inductor, diode, transistor, and two capacitors, all placed on a $Si$ panel. The geometry and material characteristics of both the components and the interposer are detailed in Tables~\ref{table: Table_InterposerGeo} and \ref{table: Table_InterposerMat}. 

The optimization process must be conducted within a bounded yet vast design space, where each component can be placed anywhere on the $Si$ panel of the interposer such that the final interposer assembly achieves minimum temperature and stress levels while occupying a minimum overall footprint.
Like the ASIC, the interposer includes both electrical and mechanical components; however, this study focuses on the thermomechanical design aspect, that is on the overall thermoelastic response given the placement of each electrical component which behaves as a thermal source. The primary objective is to design an optimal interposer assembly that satisfies specific thermoelastic constraints.
Unlike the ASIC problem, which was addressed in a discrete design space, the design space for the interposer is continuous, therefore allowing components to be placed anywhere within the $Si$ panel (see Fig.~\ref{fig: Inter_Geo}(a)). In addition, the optimization process must avoid overlapping elements, hence ensuring that components of different sizes do not interfere with one another within the interposer.

\begin{table}[ht]
    \centering
    \begin{tabular}{|c|c|c|c|c|}
        \cline{1-5}  
        \multicolumn{1}{|c}{\textbf{Component}} & \multicolumn{1}{|c|}{\textbf{Dimensions}} & \multicolumn{1}{c|}{\textbf{Power dissipation}} & \multicolumn{1}{c|}{\textbf{Material}} & \multicolumn{1}{c|}{\textbf{Interface}} \\ 
        \multicolumn{1}{|c}{} & \multicolumn{1}{|c|}{\textbf{($mm \times mm \times mm$)}} & \multicolumn{1}{c|}{\textbf{($mW$)}} & \multicolumn{1}{c|}{} & \multicolumn{1}{c|}{\textbf{material}} \\ \cline{1-5} 
        \hline
        \hline
		Interposer & $15.0 \times 15.0 \times 0.725$ & - & $Si$ & -\\
        $Si$ panel &  &  &  & \\
        \hline
        Inductor ($1$) & $4.9 \times 4.9 \times 4.0$ & 10 & $Cu$ & $Au$\\
        \hline
        Feedback & $4.3 \times 2.6 \times 0.2$ & 850 & $Si$ & $Au$\\
        controller chip ($2$) &  &  &  & \\
        \hline
        Capacitor 1 ($3$) & $2.0 \times 1.25 \times 1.25$ & 10 & Glass & $Au$\\
        \hline 
        Capacitor 2 ($4$) & $2.0 \times 1.25 \times 1.25$ & 10 & Glass & $Au$\\
        \hline 
        ASIC ($5$) & $2.0 \times 2.0 \times 0.35$ & 415 & $Si$ & $Au$\\
        \hline
        Transistor ($6$) & $1.7 \times 0.92 \times 0.685$ & 500 & $GaN$ & $Au$\\
        \hline
        Diode ($7$) & $1.0 \times 1.0 \times 0.15$ & 300 & $GaN$ & $Au$\\
        \hline
    \end{tabular}
    \caption{Summary of the geometric parameters and materials used in the FE model of the HI interposer. Each component is indicated by its corresponding number in Fig.~\ref{fig: Inter_Geo}(a).}
    \label{table: Table_InterposerGeo}
\end{table}

\begin{table}[ht]
    \centering
    \begin{tabular}{|c|c|c|c|c|c|}
        \cline{1-6}  
        \multicolumn{1}{|c}{\textbf{Material}} & \multicolumn{1}{|c|}{\textbf{Density}} & \multicolumn{1}{c|}{\textbf{Young's}} & \multicolumn{1}{c|}{\textbf{Poisson's}} & \multicolumn{1}{c|}{\textbf{Thermal}} & \multicolumn{1}{c|}{\textbf{Coeff. of thermal}} \\ 
        \multicolumn{1}{|c}{} & \multicolumn{1}{|c|}{$\rho$ ($\frac{kg}{m^3}$)} & \multicolumn{1}{c|}{\textbf{modulus}} & \multicolumn{1}{c|}{\textbf{ratio}} & \multicolumn{1}{c|}{\textbf{conductivity}} & \multicolumn{1}{c|}{\textbf{expansion}} \\
        \multicolumn{1}{|c}{} & \multicolumn{1}{|c|}{} & \multicolumn{1}{c|}{$E~(GPa)$} & \multicolumn{1}{c|}{$\nu$} & \multicolumn{1}{c|}{$\kappa~(\frac{W}{mK})$} & \multicolumn{1}{c|}{\textbf{} $\alpha_v~(\frac{1}{K})$} \\ \cline{1-6} 
        \hline
        \hline
		$Si$ & 2329 & 170 & 0.28 & 131 & 2.6e-6\\
        \hline
        $Cu$ & 8960 & 110 & 0.35 & 400 & 17e-6\\
        \hline
        Glass & 2210 & 70 & 0.22 & 1.4 & 7e-7\\
        \hline
        $GaN$ & 6070 & 295 & 0.25 & 130 & 5.59e-6\\
        \hline
        $Au$ & 19300 & 49 & 0.44 & 317 & 14.2e-6\\
        \hline
    \end{tabular}
    \caption{Material properties used in the FE model of the HI interposer.}
    \label{table: Table_InterposerMat}
\end{table}

\subsection{Thermoelastic FE model of the interposer}
\label{ssec: FE_interposer}

\begin{figure}[h!]
	\centering
	\includegraphics[width=1.0\linewidth]{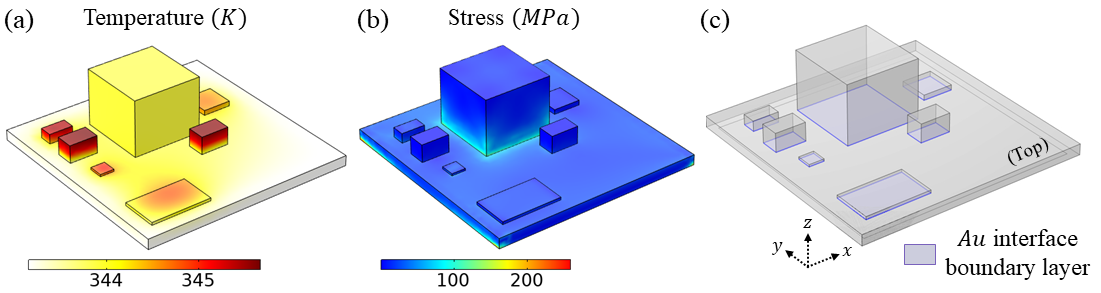}
	\caption{Numerical results obtained via FE simulations and showing (a) the temperature profile, (b) the von Mises stress profile. (c) The highlighted parts (light blue shade) represent the $Au$ interface boundary layer connecting the components to the $Si$ panel.}
	\label{fig: Inter_FESim}
\end{figure}

Developing a forward solver to evaluate the physical response of the interposer for all possible configurations defined during the optimization process is an essential aspect of design optimization under physical constraints. To analyze the thermoelastic response of the interposer, we develop a 3D HI interposer model using the finite element (FE) software package COMSOL Multiphysics$\textsuperscript{\textregistered}$. The geometry and the corresponding mesh are shown in Fig.~\ref{fig: Inter_Geo}(a) and (b). The corresponding geometric and material parameters are provided in Tables~\ref{table: Table_InterposerGeo} and \ref{table: Table_InterposerMat}. 

The forward model for the interposer is set up as a stationary thermoelastic simulation to solve Eqs.\ref{eqn: GE_thermal}-\ref{eqn: BC_thermoelasticity} introduced in \S\ref{ssec: Theory_TE}. Adiabatic thermal boundary conditions are applied to all surfaces except for a thin square annulus of thickness $t_i=0.5~mm$ on the bottom surface of the $Si$ panel enforced with isothermal boundary conditions (at $T=343.15~K$), as shown in Fig.~\ref{fig: Inter_Geo}(c). Additionally, each of the seven components serves as a thermal heat source with power dissipation values indicated in Table~\ref{table: Table_InterposerGeo}. Mechanically, the interposer geometry is constrained through fixed boundary conditions on the lower edges of the $Si$ panel (see Fig.~\ref{fig: Inter_Geo}(c)). Further, the developed model is meshed with tetrahedral elements (see Fig.~\ref{fig: Inter_Geo}(b)) with a minimum element size of $\Delta h =50~\mu m$. The FE model is solved under a range of random combinations of different component placements and the corresponding maximum temperature ($T'_m$) and maximum stress ($\sigma'_m$) distributions on the top surface of the $Si$ panel are recorded for each simulation. Specifically, $\sigma'_m$ refers to the maximum \textit{von Mises stress} in the design. Fig.~\ref{fig: Inter_FESim}(a) and Fig.~\ref{fig: Inter_FESim}(b) show the temperature profile and von Mises stress profile, respectively, for a sample interposer assembly evaluated using FE simulations. 

Before moving ahead with the implementation of the optimization process, it is essential to outline the modeling assumptions. Two key modeling approximations have been made to simplify the complexities of the FE solver: 1) The thermal and mechanical properties of each component in the assembly are determined by the characteristics of the material with the highest volume fraction within that component. In other words, the components are considered as homogenized units, with the material properties assigned based on the predominant material by volume as indicated in Table \ref{table: Table_InterposerMat}. 2) The components are connected to the interposer $Si$ panel via micron scale $Au$ solder joints (or interconnects, as in the case of ASIC). However, since simulating the interposer with a large number of microscale solder joints would be computationally intensive due to its multiscale nature, we simplify this by modeling an interface boundary layer of $Au$ between each component and the interposer $Si$ panel to emulate the solder bump connections, as shown in Fig.~\ref{fig: Inter_FESim}(c). These assumptions enable us to create a forward model capable of effectively simulating multiple interposer designs.
From Table \ref{table: Table_InterposerGeo}, we notice that the ASIC is also a component on the HI interposer assembly. However, in the interposer design, the ASIC is treated as a homogenized unit with a $Au$ boundary layer rather than an array of interconnects. While this approximation serves as the preliminary model for the HI interposer design, future studies can incorporate a detailed ASIC model within the interposer assembly in order to simulate more accurately the multiphysics response.


However, integrating a FE forward solver directly into the optimization process significantly increases computational time. Therefore, developing a faster and accurate surrogate model is essential. The following section highlights the use of DNN to build such a surrogate model. While the DNN model accelerates the process, it still relies on the FE model introduced here to train an accurate data-driven surrogate model.

\subsection{Proposed deep RL methodology}

\begin{figure}[h!]
	\centering
	\includegraphics[width=1.0\linewidth]{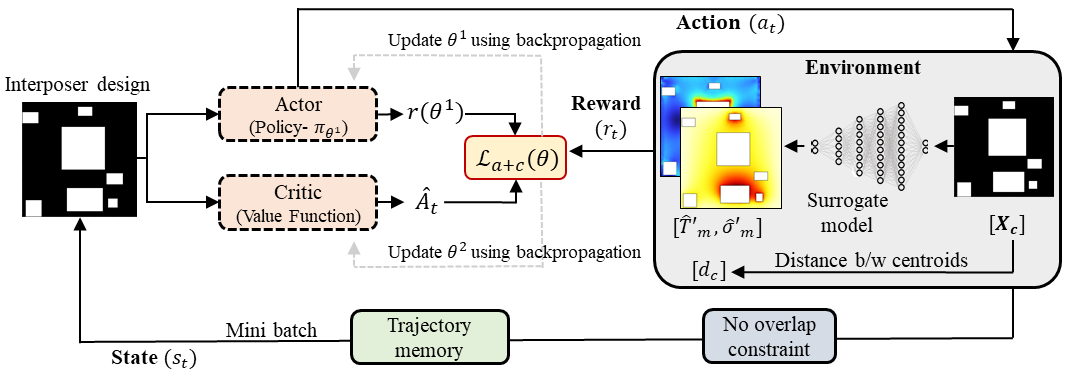}
	\caption{Schematic illustrating the proposed deep RL algorithm for interposer design. The proximal policy optimization (PPO) with actor-critic architecture is highlighted.}
	\label{fig: RL_Inter}
\end{figure}

We propose a deep reinforcement learning (RL) framework to optimize the design of a HI interposer assembly as illustrated in Fig.~\ref{fig: RL_Inter}. This approach leverages deep learning integrated with RL to estimate the optimal design that achieves the desired thermoelastic behavior for a tightly packed interposer. The ASIC interconnect design discussed in \S\ref{ssec: ASIC_RLModel} dealt with only two design variables and well-defined discrete bounds. This problem formulation allowed representing and evaluating the values for each $s_t-a_t$ pair in a grid search form without complex value approximators.
However, given the various combinations of possible placements for the seven components, the interposer design problem introduces a significantly larger design space with numerous possible actions. Unlike the basic RL models like Q-learning, the tabular or grid representation of values becomes impractical when dealing with such high-dimensional and continuous $s_t-a_t$ pairs.

In a deep RL framework, the agent selects an optimal action $a_t$ for the current state $s_t$ based on the policy $\pi_{\theta}$ to determine the design and interact with the environment (i.e. the forward solver). While the policy $\pi$ in Q-learning was implicitly derived from the $Q$ values, in deep RL models the policy $\pi_{\theta}$ is learned using deep neural networks, with the policy being parameterized by the network weights $\theta$. Beginning with an initial guess of a design, similar to the basic RL models, the deep RL algorithm also iteratively searches for the optimal design using the $r_t$ estimated from the physical responses calculated by a forward solver. The agent continues exploring the design space until the cumulative $r_t$ is maximized.

The forward solver and the deep RL algorithm are the two key components of the proposed deep RL framework. The first part of this section will introduce the forward solver embedded within the RL model, which is crucial for calculating $r_t$ for optimal design parameters. The second part will elaborate on the details of the deep learning integrated RL algorithm used and the optimization process followed.

\subsubsection{DNN based surrogate model for forward solver}
\label{ssec: Inter_SurrogateModel}

While existing optimization approaches commonly employ traditional forward solvers, such as FE solver, integrating a traditional solver significantly increases the computational time for the design optimization process. To address this issue, we develop a surrogate forward solver using a deep neural network (DNN). Specifically, we design a feed-forward DNN that is trained to accurately predict $\hat{T}'_m$ and $\hat{\sigma}'_m$ using as input the centroid positions of all the seven components, as shown in Fig.~\ref{fig: RL_Inter}. Here, the quantities with $\hat{(.)}$ represent the surrogate model predictions. The surrogate model is trained across a range of interposer designs with varying placements of the components. Once trained, we leverage the interpolation capability of the DNN to embed the pre-trained model as a surrogate forward solver within the deep RL framework (see Fig.~\ref{fig: RL_Inter}).

\begin{figure}[h!]
	\centering
	\includegraphics[width=1.0\linewidth]{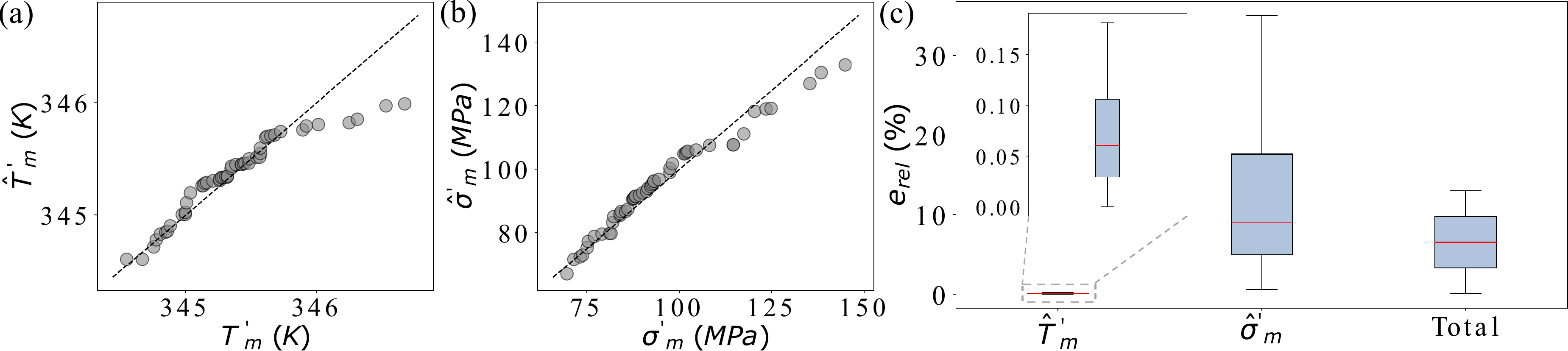}
	\caption{Plots comparing the prediction of the trained surrogate model with the ground truth based on the test dataset for (a) temperature ($T'_m$) and (b) stress ($\sigma'_m$). The closer the predictions are to the dashed line the better the accuracy of the surrogate model. (c) Box plot highlighting the relative error $e_{rel}$ (in $\%$) distribution across the test dataset for $T'_m$, $\sigma'_m$, and $T'_m -\sigma'_m$ together (total). The red line in each boxplot represents the median $e_{rel}$.}
	\label{fig: InterSurr_data}
\end{figure}

The surrogate model is a fully connected neural network architecture that consists of four hidden layers, each with 30 neurons, followed by $ReLU$ activation function in each layer. Moreover, the training data for the surrogate DNN model is generated using the FE model presented in \S\ref{ssec: FE_interposer}. The FE model is solved for 1000 different physically plausible models (ensuring no component overlaps), with varying placements of the seven components. After simulating all the interposer designs, the centroid positions of all components in each interposer design are recorded as the network input data $\textbf{X}_c=[\{x^0_c, y^0_c\}, \{x^1_c, y^1_c\},... \{x^6_c, y^6_c\}]$, while the corresponding $T'_m$ and $\sigma'_m$ are recorded as the output data. Using this labeled input-output dataset, the network is trained to find the optimal network parameters $\theta$ by minimizing the mean square error (MSE) loss function as follows
\begin{equation}
\begin{split}
    \mathcal{L}'(\theta) =&~ \omega_1 \mathcal{L}'_T(\theta) + \omega_2 \mathcal{L}'_{\sigma}(\theta) \\
    =&~ \frac{\omega_1}{N} \sum^N_{i=1} \Big( \hat{T}'_m(\theta) - T'_m \Big)^2 + \frac{\omega_2}{N} \sum^N_{i=1} \Big( \hat{\sigma}'_m(\theta) - \sigma'_m \Big)^2
\end{split}
\end{equation}
where $\mathcal{L}'_T$ is the MSE loss for temperature and $\mathcal{L}'_{\sigma}$ is the MSE loss for stress. In addition, the network is trained with $N$ training data samples with $\omega_1=1$ and $\omega_2 =1$ as the weighting factors for $\mathcal{L}'_T$ and $\mathcal{L}'_{\sigma}$, respectively.

Out of 1000 data samples, $N + N_{val} = 950$ samples are split into training and validation datasets in a 75:25 ratio. The remaining $N_{test} = 50$ samples are used as the test dataset. This DNN is trained using Adam optimizer with a learning rate of 5e-4, weight decay of 1e-2, and batch size of 64 samples for 1000 epochs.

The performance of the trained surrogate model is evaluated qualitatively by plotting the comparison between the model prediction and the ground truth for temperature and stress in Fig.~\ref{fig: InterSurr_data}(a) and (b), respectively. The closer the predictions are to the dashed straight lines, the higher the chances of the prediction and ground truth belonging to the same distribution. This comparison indicates low variability and high accuracy in the network prediction. In addition, this is corroborated quantitatively with a box plot in Fig.~\ref{fig: InterSurr_data}(c). The box plot highlights the variation of relative $\%$ error ($e_{rel}$) in prediction of the temperature, stress, and temperature-stress together (total). Moreover, calculating the mean prediction error across the test dataset, evaluates $\bar{e}_{rel} |_{T'_m} = 0.074 \%$, $\bar{e}_{rel} |_{\sigma'_m} = 12.96 \%$, and $\bar{e}_{rel} |_{T'_m-\sigma'_m} = 6.52 \%$. While the overall relative error $\bar{e}_{rel} |_{T'_m-\sigma'_m}$ across the test dataset indicates high prediction accuracy, the relative errors in temperature and stress predictions vary significantly. This variation can be attributed to the differences in ranges of temperature and stress in the training data. While temperature varies by less than $2~K$ for the proposed interposer designs, the variation in stress for the same models ranges around $100~MPa$. Despite data normalization during training, smaller variations in temperature leads to higher prediction accuracy as indicated by smaller $\bar{e}_{rel} |_{T'_m}$, whereas larger variation in stress results in slightly higher $\bar{e}_{rel} |_{\sigma'_m}$. However, the low overall prediction error $\bar{e}_{rel} |_{T'_m-\sigma'_m}$ highlights the reliable prediction accuracy of the proposed surrogate model. The surrogate DNN model is trained using Python 3.8 in Pytorch API on NVIDIA A100 Tensor Core GPU with 80GB memory. 

\subsubsection{Proximal policy optimization (PPO) deep RL algorithm}

We choose to implement a policy gradient based proximal policy optimization (PPO) deep RL algorithm to address the interposer design optimization problem. The policy gradient methods optimize a policy (i.e. a strategy to choose an action) by adjusting the policy parameters to maximize the expected cumulative reward. This approach differs from value based methods like Q-learning, which estimates the value function and implicitly derives the policy from it. 
In this study, we develop a PPO algorithm with an \textit{actor-critic} deep neural network architecture as shown in Fig.~\ref{fig: RL_Inter}. While the PPO ensures stable iterative updates by clipping the probability ratio between updated and new policies during training, the actor-critic architecture combines the advantages of policy gradient and value function estimation approaches. The actor network selects actions based on the current policy, while the critic network estimates a value function that assesses the quality of the selected action for the given state, as depicted in Fig.~\ref{fig: ActorCritic_NN}.

Let us consider a policy network $\pi_{\theta^1}$, which is optimized through the neural network parameters $\theta^1$. The network takes the current state $s_t$ as its input and outputs an action $a_t$. More specifically, in the context of the interposer problem, the spatial configuration of the $Si$ layer is provided as input $s_t$, while the policy network (actor) is trained to output an action $a_t$, representing the placement location of each component. For a continuous action space in the interposer design, the actor network predicts a probability distribution for placing a component in $s_t$. Then, the $a_t$ is randomly sampled from this predicted distribution to increase the exploration of the design space. As training progresses, the actor network learns to optimize the policy $\pi_{\theta^1}$, leading to the identification of the optimal component placement.
In order to find the optimal policy $\pi_{\theta^1}$, a policy gradient approach estimates the gradient of $\pi_{\theta^1}$ and then applies the gradient ascent algorithm. For a probability ratio denoted $r(\theta^1)=\frac{\pi_{\theta^1} (a_t | s_t)}{\pi_{\theta^1_{old}} (a_t | s_t)}$, such that $r(\theta^1_{old})=1$, the regular policy gradient approach, called trust region policy optimization, updates the policy by maximizing an objective \cite{schulman2015trust} as follows
\begin{equation}
\label{eqn: TRPO_obj}
    \mathcal{L}'_a(\theta^1) = \mathbb{\hat{E}}_t\ \left[ \frac{\pi_{\theta^1} (a_t | s_t)}{\pi_{\theta^1_{old}} (a_t | s_t)} \hat{A}_t \right] = \mathbb{\hat{E}}_t\ \left[ r(\theta^1) \hat{A}_t \right]
\end{equation}
where $\pi_{\theta^1}$ represents the policy with updated parameters $\theta^1$, $\pi_{\theta^1_{old}}$ represents the policy parameters $\theta^1_{old}$ before the update, and $\hat{A}_t$ is an estimator of the advantage function at time step $t$. Here, $\mathbb{\hat{E}}_t[.]$ represents the expectation that indicates the empirical mean over a finite batch of training samples. The objective function in Eq.~\ref{eqn: TRPO_obj} demonstrates a conservative policy iteration, but it has a limitation; maximizing the $\mathcal{L}'_a$ could result in an excessively large policy update with a solution well beyond the design space. To address this concern, the PPO algorithm \cite{schulman2017proximal} introduces a clipped objective function that penalizes large changes as follows
\begin{equation}
\label{eqn: PPO_obj}
    \mathcal{L}_a(\theta^1) = \mathbb{\hat{E}}_t\ \left[ \min \Big( r(\theta^1) \hat{A}_t, \text{clip} \left( r(\theta^1), 1-\mathcal{E}, 1+\mathcal{E} \right)\hat{A}_t \Big) \right]
\end{equation}
where $\mathcal{E}=0.2$ depicts the clipping bounds. While the first term inside $\min(.)$ is the same as Eq.~\ref{eqn: TRPO_obj}, the second term modifies the objective by clipping the probability ratio to the region $[1-\mathcal{E}, 1 + \mathcal{E}]$. Finally, $\mathcal{L}_a$ takes the minimum of the clipped and unclipped objectives so that the final objective is a lower bound on the unclipped objective. Thus, $\mathcal{L}_a$ (Eq.~\ref{eqn: PPO_obj}) is the objective function for the actor network to find the optimal policy $\pi_{\theta^1}$ by optimizing the network parameters $\theta^1$.

The primary challenge in policy gradient methods lies in reducing the variance of the gradient estimates such that consistent progress towards better policies can be made \cite{schulman2017proximal}. The introduction of an actor-critic architecture makes a significant impact in this regard by introducing advantage function ($\hat{A}_t$) estimators that make use of $s_t$ value function $V(s_t)$. The value function is estimated using another neural network called the critic network. While the actor network learns the optimal policy and generates actions, the critic network is responsible for estimating the value corresponding to the predicted action. To this end, the critic network based value estimator is trained to satisfy the following squared error loss
\begin{equation}
    \label{eqn: critic}
    \mathcal{L}_c(\theta^2) = \mathbb{\hat{E}}_t \left[ \left( \hat{V}_{\theta^2}(s_t) - V(s_t) \right)^2 \right]
\end{equation}
where $\theta^2$ is the critic network parameters, $\hat{V}_{\theta^2}(s_t)$ is the predicted values, and $V(s_t)$ is the target value. Note that the target value is empirically estimated as $V(s_t)= \hat{A}_t + V_{old}(s_t)$, where $V_{old}(s_t)$ is the previous value in memory for $s_t$.

Using the objective functions presented above, the actor-critic architecture is trained together by combining Eqs.~\ref{eqn: PPO_obj} and \ref{eqn: critic} to maximize the following objective in each iteration
\begin{equation}
    \label{eqn: actor_critic}
    \mathcal{L}_{a+c}(\mathrm{\theta}) = \mathbb{\hat{E}}_t \left[ \mathcal{L}_a(\theta^1) - c_1 \mathcal{L}_c(\theta^2) \right]
\end{equation}
where $c_1=0.5$ is the coefficient and $ \mathbf{\theta} = [\theta^1, \theta^2]$. Further, the advantage function, integral to the evaluation of both $\mathcal{L}_a$ and $\mathcal{L}_c$, is estimated for $T$ timesteps
\begin{equation}
\begin{split}
    \label{eqn: advantage}
    \hat{A}_t =& \delta_t + (\gamma \lambda)\delta_{t+1} + ...(\gamma \lambda)^{T-t+1}\delta_{T-1}, \\
    s.t. &~~~~\delta_t = r_t + \gamma V(s_{t+1}) - V(s_t)
\end{split}
\end{equation}
where $t$ is timestep in the range $[0, T]$, $r_t$ is the reward function, $\gamma=0.99$ is the discount factor, and $\lambda=0.95$ is the smoothing parameter. In this study, $r_t$ is defined as a function of $\hat{T}'_m$ and $\hat{\sigma}'_m$ as follows
\begin{equation}
    \label{eqn: r_interposer}
    r_t = \frac{\omega_1}{\hat{T}'_{m_n}} + \frac{\omega_2}{\hat{\sigma}'_{m_n}} + \frac{\omega_3}{d_{c_n}}
\end{equation}
where, $\hat{T}'_{m_n}$ is the normalized maximum temperature ($\hat{T}'_m$), $\hat{\sigma}'_{m_n}$ is the normalized maximum stress ($\hat{\sigma}'_m$), $d_{c_n}$ is the normalized distance between centroids of the components ($d_c$), and $\{\omega_1, \omega_2, \omega_3 \}$ are the weighting factors. Note that, the $\hat{T}'_m$ and $\hat{\sigma}'_m$ for each interposer design configuration is evaluated using the trained surrogate model (\S\ref{ssec: Inter_SurrogateModel}).

In summary, the proposed PPO based RL model with actor-critic networks, is trained to maximize the objective in Eq.~\ref{eqn: actor_critic}. The result is a maximization of the cumulative reward $r_t$, in order to determine the optimal placement of components on the interposer such that the resulting optimal configuration satisfies minimum temperature, stress, and footprint constraints.

\subsubsection{Actor-critic neural network architecture}

\begin{figure}[h!]
	\centering
	\includegraphics[width=1.0\linewidth]{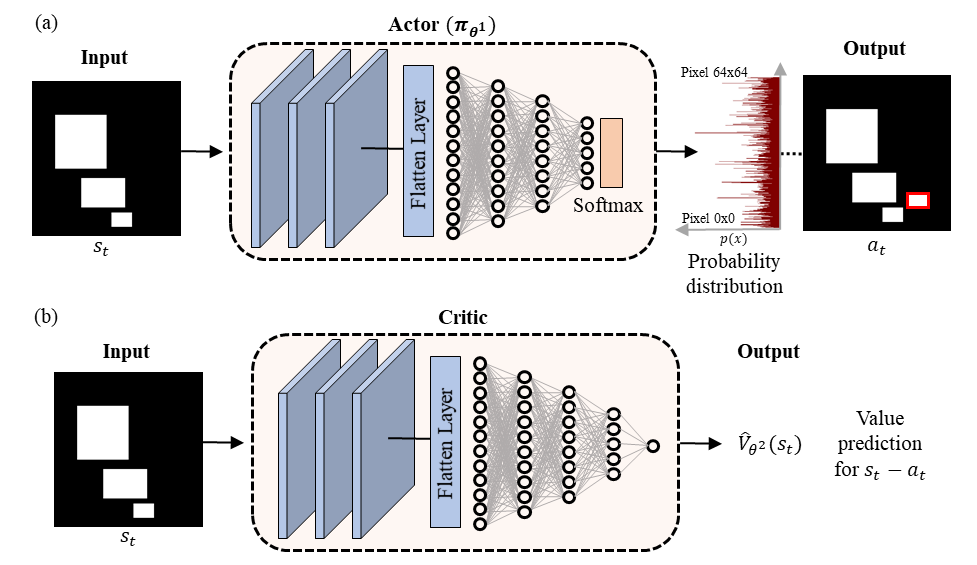}
	\caption{Schematic illustration of the neural network architectures of (a) the actor and (b) the critic. The inputs and outputs of both networks are also highlighted. Additionally, the illustration depicts a design iteration example where three components have already been placed, forming the current input state $s_t$. The actor network is tasked with predicting $a_t$, which identifies the pixel position for placing the fourth component (highlighted by a red box). Simultaneously, the critic network is responsible for predicting the value corresponding to this $s_t-a_t$ pair.}
	\label{fig: ActorCritic_NN}
\end{figure}

In the previous section, we briefly introduced the roles and functions of the actor and critic neural networks within the proposed deep RL framework. Here, we will elaborate on the architecture of these networks and discuss their implementation in detail.

The \textit{actor} is a deep neural network with network parameters $\theta^1$ as illustrated in Fig.~\ref{fig: ActorCritic_NN}(a). It takes the current state $s_t$ of the interposer design as input and predicts a probability distribution over possible actions $a_t$ within the design space as output. Specifically, for the interposer design, the 2D design space of the $Si$ layer is represented as a binary image input. In an input image of size $64 \times 64$ pixels, the $Si$ layer without components is represented with pixel value of 0 (black), while each placed component is represented by pixel value of 255 (white). The actor network consists of three 2D convolutional layers with zero padding and 64 channels each. These convolutional layers use a $3 \times 3$ kernel for all convolution operations. Following the convolutional layers, the network includes three fully connected linear layers. In addition, all the neural layers are followed by the $ReLU$ activation function. The final linear layer is followed by a softmax layer, which produces a probability distribution output of size $(64*64) \times 1$, corresponding to the possible actions across all input pixels.

The \textit{critic} is a second deep neural network with network parameters $\theta^2$ as shown in Fig.~\ref{fig: ActorCritic_NN}(b). It takes the current state $s_t$ of the interposer design as input and predicts a value $\hat{V}_{\theta^2}$ as the output for that state. Similar to the actor network, the critic receives the 2D design space of the $Si$ layer as a binary image input. The overall network architecture of the critic is similar to the actor. It consists of three 2D convolutional layers, each with zero padding and 64 channels. These are followed by three fully connected linear layers. Additionally, all the neural layers are followed by the $ReLU$ activation function, and all convolution operations are performed using a $3 \times 3$ kernel. However, unlike the actor network, which outputs a probability distribution, the critic outputs a single value. Therefore, the final layer of the critic is a linear layer of size $1 \times 1$.

Both the actor and the critic networks are trained jointly within the RL framework to find the optimal design. This is achieved by optimizing the parameters $\theta=[\theta^1, \theta^2]$ by maximizing Eq.~\ref{eqn: actor_critic}. Alternatively, this can be represented as a minimization problem as follows 
\begin{equation}
\label{eqn: theta_argmin}
    \theta^* = \arg \min_{\theta} - \mathcal{L}_{a+c} (\theta)
\end{equation}
It is important to note that, while the RL model aims to maximize the objective through gradient ascent, in practice we leverage the automatic differentiation capabilities of neural networks by minimizing the objective function (Eq.~\ref{eqn: theta_argmin}) using gradient descent through backpropagation. The deep RL framework with actor-critic DNNs is trained using Adam optimizer with a learning rate of 3e-4, batch size of 10 samples for 100 epochs, and implemented in Python 3.8 using Pytorch API on NVIDIA A100 Tensor Core GPU with 80GB memory. Unlike the surrogate model, the actor-critic networks are not pre-trained offline. Instead, they are trained online as part of the deep RL framework, following the PPO algorithm.

\subsection{Numerical experiments and discussions}
\label{ssec: Interposer_results}

In this section, we evaluate the performance of the proposed deep RL framework by assessing its ability to solve for optimal HI interposer designs under thermoelastic and geometric constraints. First, we analyze the optimization process employed by the proposed model. Then, we investigate the optimal designs obtained under different thermoelastic and geometry constraints and validate the results using FE based thermoelastic simulations.

\subsubsection{Placement optimization of components on the interposer}

\begin{figure}[h!]
	\centering
	\includegraphics[width=1.0\linewidth]{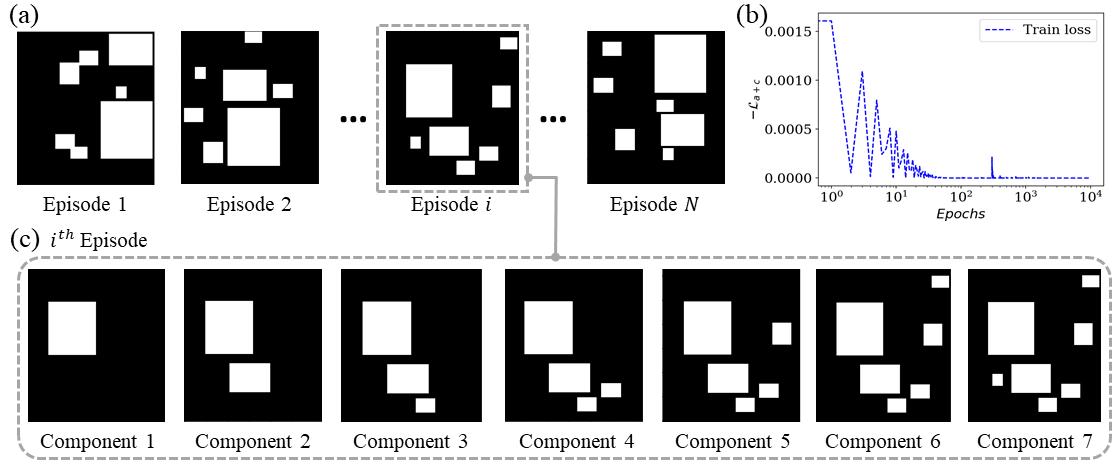}
	\caption{Evolution of the interposer configuration during the design optimization process. (a) The deep RL design optimization process unfolds over $N$ episodes. (b) The loss plot demonstrates the minimization of the total loss with training epochs. (c) Every episode of the learning process includes seven iterative stages for accurate placement of the seven components within the design space.}
	\label{fig: deepRL_evolution}
\end{figure}

The design problem aims to find an optimal interposer configuration such that the resulting placement of the components minimizes $T'_m$, $\sigma'_m$, and $d_c$ on the $Si$ panel. An example of this scenario is illustrated in Fig.~\ref{fig: deepRL_evolution}(a) obtained from the maximization of $r_t$ in Eq.~\ref{eqn: r_interposer}. This reward maximization is directly associated to the maximization of $\mathcal{L}_{a+c}$ in Eq.~\ref{eqn: actor_critic}. In other words, the actor-critic models are trained to enforce loss minimization of $-\mathcal{L}_{a+c}$ as shown in Fig.~\ref{fig: deepRL_evolution}(b). 

The design process involves two stages. First, the deep RL model is trained over $N=100$ episodes to identify the optimal placement of all seven components on the interposer, as shown in Fig.~\ref{fig: deepRL_evolution}(a). Second, each episode consists of seven iterations, corresponding to the sequential placement of the seven components in order of decreasing surface area, marked as components 1-7 in Fig.~\ref{fig: deepRL_evolution}(c).
Another crucial aspect of the design process is to avoid any overlap between components on the interposer (see Fig.~\ref{fig: RL_Inter}). This \enquote{no-overlap} condition is implemented as a hard-coded constraint, ensuring that each component is placed without overlapping with any previously placed components. In the case of a chosen $a_t$ leading to a component overlap, the iteration is repeated until an action $a_t$ satisfies the no-overlap condition to maintain the integrity of the interposer design.

After every episode, the surrogate model (forward solver) evaluates the thermoelastic response $T'_m$ and $\sigma'_m$ for the complete configuration. However, since placing individual components does not immediately contribute to the rewards associated with $T'_m$ and $\sigma'_m$, this situation results in a sparse reward problem. The distance contribution $d_c$ is evaluated after the placement of each component, providing a marginal contribution to $r_t$ in each iteration. In other words, significant reward contributions from $T'_m$ and $\sigma'_m$ are only added after the completion of each episode, making the thermoelastic constraints sparsely enforced while geometric constraints are continuously enforced at every iteration.

\subsubsection{Optimal interposer design under thermoelastic and geometric constraints}

\begin{figure}[h!]
	\centering
	\includegraphics[width=1.0\linewidth]{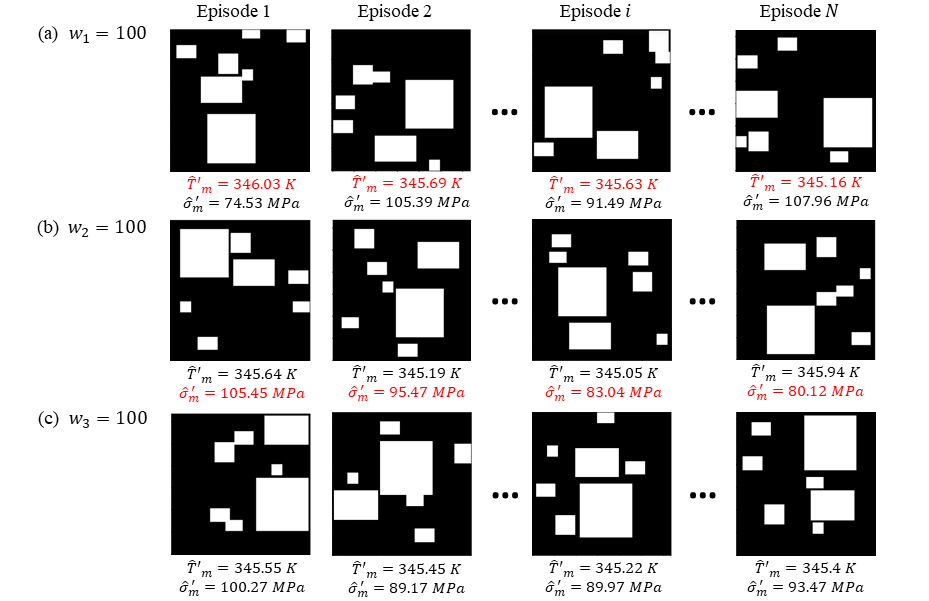}
	\caption{The evolution of the optimal design of the interposer for (a) higher weighting factor for $\hat{T}'_m$ - $\{ \omega_1=100, \omega_2=1, \omega_3=1 \}$, (b) higher weighting factor for $\hat{\sigma}'_m$ - $\{ \omega_1=1, \omega_2=100, \omega_3=1 \}$, and (c) higher weighting factor for $d_c$ - $\{ \omega_1=1, \omega_2=1, \omega_3=100 \}$.}
	\label{fig: deepRL_OptimalDesign}
\end{figure}

This section explores the performance of the proposed deep RL model in terms of the optimal design of the interposer. Given that the study employs a multi-objective reward function (Eq.~\ref{eqn: r_interposer}) incorporating temperature, stress, and geometric constraints, the weighting factor ($\omega_i$, where $i=\{1,2,3 \}$) has a critical effect on the optimal design. In other words, the optimal design parameters may vary depending on the chosen $\omega_i$ in Eq.~\ref{eqn: r_interposer}. Therefore, it is important to analyze the performance of the optimization model across different weighting factors as illustrated in Fig.~\ref{fig: deepRL_OptimalDesign}.
Finally, the RL design predictions are validated using thermoelastic FE simulations.

We specifically analyze the performance of the deep RL model under varying weighting factors conditions: one emphasizing temperature, one focusing on stress, and one targeting centroid distances. Fig.~\ref{fig: deepRL_OptimalDesign}(a) illustrates the evolution stages for a case with higher contribution from the temperature weighting factor as $\{ \omega_1=100, \omega_2=1, \omega_3=1 \}$. As described earlier, by the $N^{th}$ episode, the actor-critic architecture learns to predict optimal design state $s^*_t$. The results show a noticeable reduction in temperature values across episodes, demonstrating the influence of the higher weighting factor $\omega=100$ in Eq.~\ref{eqn: r_interposer}, which guides the model to predict an interposer configuration with minimized temperature levels. The optimal configuration is validated by evaluating $T'_m$ with results from FE simulations that, once compared with the surrogate model prediction $\hat{T}'_m$, indicate less than $0.1\%$ prediction error.    

Similarly, Fig.~\ref{fig: deepRL_OptimalDesign}(b) show the design evolution for a scenario with a higher contribution from the stress term, which is obtained by setting the weighting factors as $\{ \omega_1=1, \omega_2=100, \omega_3=1 \}$. Results indicate a noticeable reduction in stress values across episodes, highlighting the effectiveness of the higher weighting factor $\omega_2=100$ in Eq.~\ref{eqn: r_interposer}. This set of weighting factors helps the model to learn and predict an interposer configuration with minimal stress levels over time. Further, the optimal configuration is validated by evaluating $\sigma'_m$ using FE simulations, where the surrogate model prediction $\hat{\sigma}'_m$ has less than $7\%$ prediction error. Note that this error is higher compared to the temperature prediction error, which can be attributed to the greater variance in stress distribution across designs compared to temperature. Consequently, the trained networks achieve higher accuracy in predicting temperature.
Finally, Fig.~\ref{fig: deepRL_OptimalDesign}(c) shows the design evolution for higher contributions of the centroid distance weighting factor, obtained by the following set of weights $\{ \omega_1=1, \omega_2=1, \omega_3=100 \}$. Unlike the improvements seen with temperature and stress, there is no significant quantitative reduction in the footprint. This behavior is due to the influence of temperature and stress contributions to the total reward function, which prevents complete minimization of the distance measure. However, the numerical results clearly show that the proposed RL framework is capable of optimizing the interposer design under multiphysics constraints.

\begin{figure}[h!]
	\centering
	\includegraphics[width=1.0\linewidth]{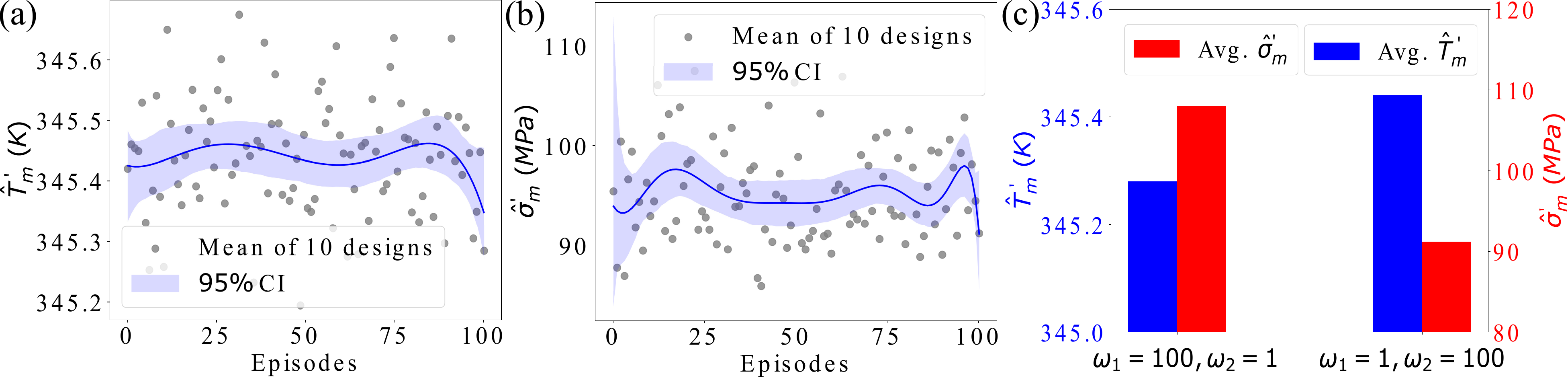}
	\caption{Schematic illustrating (a) the mean variation of $\hat{T}_m'$ with episodes, and (b) the mean variation of $\hat{\sigma}_m'$ with episodes. The plots highlight 95\% confidence interval (CI) of the predictions and each circular marker represents an interposer design averaged over ten design optimizations for each episode. (c) A bar plot highlighting the $\hat{T}_m'$ and $\hat{\sigma}_m'$ values averaged over ten design optimizations for weighting factors $\{ \omega_1=100, \omega_2=1, \omega_3=1 \}$ and $\{ \omega_1=1, \omega_2=100, \omega_3=1 \}$.}
	\label{fig: Inter_TSBar}
\end{figure}

While Fig.~\ref{fig: deepRL_OptimalDesign}(a)-(c) showcases a single interposer design predicted by the proposed deep RL model for each weighting factor scenario, it is essential to note that this high-dimensional inverse interposer design problem does not yield a unique solution. Multiple design configurations can satisfy the interposer assembly problem. Alternatively, there are multiple combinations of the components that can satisfy the reward function (Eq.~\ref{eqn: r_interposer}). Therefore, in order to validate the performance and reliability of the proposed model for the interposer design, we conducted ten optimization runs for each weighting factor and assessed its average performance. Fig.~\ref{fig: Inter_TSBar}(a) shows the $\hat{T}_m'$ variation with $N=100$ episodes for a temperature dominant weighting factor $\{ \omega_1=100, \omega_2=1, \omega_3=1 \}$, while Fig.~\ref{fig: Inter_TSBar}(b) indicates the $\hat{\sigma}_m'$ variation with $N=100$ episodes for a stress dominant weighting factor $\{ \omega_1=1, \omega_2=100, \omega_3=1 \}$, both averaged over ten optimization runs. 
In both figures (Fig.~\ref{fig: Inter_TSBar}(a) and Fig.~\ref{fig: Inter_TSBar}(b)), each circular marker represents a unique interposer design. The figures illustrate that there are multiple designs capable of achieving nearly identical temperatures in the temperature-dominant case and similar stresses in the stress-dominant case.
Both figures show that the deep RL model predicts an optimal design (either with lower temperature or stress according to the chosen weighting factor), averaged over ten runs, as the training approaches $N=100$ episodes. Moreover, as the training progresses, prediction uncertainty decreases, as evidenced by the narrowing confidence intervals in the later episodes. Finally, Fig.~\ref{fig: Inter_TSBar}(c) highlights the effectiveness and control provided by the proposed model as it is capable of identifying an optimal interposer design with minimum $\hat{T}_m'$, when temperature is prioritized, and with minimum $\hat{\sigma}_m'$, when stress is prioritized. These results showcase the versatility of the model in balancing multiple objectives and effectively controlling the design parameters across multiple optimization runs.

\section{Conclusions}

This work presented a reinforcement learning (RL) based framework to perform design optimization under multiphysics constraints. Although the framework was tested considering an illustrative example of the thermoelastic response of microelectronic components, its structure is general and applicable to different classes of design problems under multiphysics constraints. More specifically, the methodology was presented via two different benchmark problems drawn from microelectronic applications. The first problem involved the optimization of the interconnect configuration for an application-specific integrated circuit (ASIC) chip, while the second focused on determining the optimal assembly of components for a heterogeneously integrated (HI) interposer. These design problems are addressed using two distinct RL models.

First, a basic RL model was introduced to address the interconnect design problem for an ASIC chip, which involved two design variables addressing the optimization of shape and placement within a discrete design space. A Q-learning based RL model was used to find the optimal configuration that minimizes temperature and stress levels in the ASIC. The RL model was designed to effectively identify the best combination of design variables while adhering to the specified thermoelastic constraints.
Second, a deep RL model was developed to address the high-dimensional placement optimization of a HI interposer within a continuous design space. By using a proximal policy optimization (PPO) algorithm with actor-critic network architecture, this deep RL model was capable of effectively determining the optimal assembly of components while minimizing the placement footprint. The method was capable of accurately satisfying both the temperature and stress constraints, hence solving the design problem under both thermoelastic and geometric constraints. 

Additionally, both RL frameworks were integrated with a surrogate model, which replaced the original finite element-based forward solver. This surrogate model was based on a pre-trained deep neural network (DNN) that takes predicted design parameters as inputs, while it outputs temperature and stress levels. Both RL models provide extensive control over the optimization objectives, allowing for the evaluation of different optimal designs just by acting on the weights of the thermoelastic and geometric constraints. Moreover, numerical results highlight the remarkable ability of RL models to accurately predict optimal designs, even when faced with complex design optimization problems under multiphysics constraints. This level of control facilitates the repeated use of the framework for customized microelectronics design. In future work, we will leverage this high level of control on the design process to investigate the ability to reduce the overall design cycle time.


\bigskip

\noindent \textbf{Competing interests}

The authors declare no competing interest.\\

\noindent \textbf{Acknowledgements}

This work was supported by the Laboratory Directed Research and Development program at Sandia National Laboratories, a multimission laboratory managed and operated by National Technology and Engineering Solutions of Sandia LLC, a wholly owned subsidiary of Honeywell International Inc. for the U.S. Department of Energy’s National Nuclear Security Administration under contract DE-NA0003525.

\bibliographystyle{unsrt} 
\bibliography{Finaldraft}

\end{document}